\documentclass[aps,prd,amsmath,amssymb,preprintnumbers,nofootinbib,a4paper,preprint]{revtex4-1}
\pdfoutput=1
\usepackage{amsthm}
\usepackage{graphicx}
\usepackage{color}
\newcommand{\beq}{\begin{eqnarray}}
\newcommand{\eeq}{\end{eqnarray}}

\newcommand{\bmp}{\noindent\begin{minipage}{16cm}}
\newcommand{\emp}{\end{minipage}\vskip 7mm} % 7mm untightened

\usepackage{dcolumn}% Align table columns on decimal point
\usepackage{bm}% bold math
\usepackage{bbm}
\usepackage{subfigure}
\usepackage{pxfonts}

\theoremstyle{definition}

\theoremstyle{plain}

\usepackage{epsfig}

\usepackage{graphicx}
\usepackage{multirow}
\usepackage[margin=5pt, font=normalsize,labelfont=bf,format=plain,justification=centerlast]{caption}

\definecolor{rossoCP3}{cmyk}{0,.88,.77,.40}

\def\lsim{\mathrel{\rlap{\lower4pt\hbox{\hskip1pt$\sim$}}
    \raise1pt\hbox{$<$}}}                % less than or approx. symbol
\def\gsim{\mathrel{\rlap{\lower4pt\hbox{\hskip1pt$\sim$}}
    \raise1pt\hbox{$>$}}}                % greater than or approx. symbol

\newcommand{\be}{\begin{eqnarray}}
\newcommand{\ee}{\end{eqnarray}}

\usepackage[%dvips,
bookmarks]{hyperref}
\usepackage{color}
\hypersetup{colorlinks, bookmarksopen, bookmarksnumbered,
citecolor=verdes, linkcolor=bluc, pdfstartview=FitH, urlcolor=rossos}

\definecolor{grigio}{cmyk}{0,0,0,0.1}
\definecolor{rosa}{cmyk}{0,0.1,0.1,0.02}
\definecolor{rosino}{cmyk}{0,0.05,0.05,0.02}
\definecolor{rosas}{cmyk}{0,0.3,0.25,0.05}
\definecolor{celeste}{cmyk}{0.1,0,0,0.02}
\definecolor{giallino}{cmyk}{0,0,0.1,0.02}
\definecolor{rosso}{cmyk}{0,1,1,0.4}
\definecolor{rossos}{cmyk}{0,1,1,0.55}
\definecolor{rossoc}{cmyk}{0,1,1,0.2}
\definecolor{blu}{cmyk}{1,1,0,0.3}
\definecolor{blus}{cmyk}{1,1,0,0.5}
\definecolor{bluc}{cmyk}{1,1,0,0.1}
\definecolor{blucc}{cmyk}{0.7,0.5,0,0}
\definecolor{viola}{cmyk}{0,1,0,0.6}
\definecolor{viola2}{cmyk}{0,1,0.2,0.6}
\definecolor{verde}{cmyk}{0.92,0,0.59,0.25}
\definecolor{verdec}{cmyk}{0.92,0,0.59,0.15}
\definecolor{verdes}{cmyk}{0.92,0,0.59,0.4}
\definecolor{verdino}{cmyk}{0.12,0,0.09,0.02}
\definecolor{giallo}{cmyk}{0,0,1,0}
\definecolor{gialloverde}{cmyk}{0.44,0,0.74,0}
\definecolor{Titolo}{rgb}{0.752941176,0.576470588,0.992156863}% #C093FD
\definecolor{altro}{rgb}{0.094117647,0.650980392,0.643137255}% #24A6A4
\definecolor{Peanuts}{rgb}{0.2, 0.4, 0.6}% #336699
\definecolor{Pean1}{rgb}{0.6, 0.8, 0.4}% #99cc66
\definecolor{BHO}{rgb}{0.2, 0.8, 1}% #33CCFF
\definecolor{Daria}{rgb}{0, 0.9412, 0}% #00F000 e' uguale al verde u_u
\definecolor{UniPi}{rgb}{0.2549, 0.4627, 0.6275}% #4176a0
\definecolor{UniPidue}{rgb}{0.3216, 0.5804, 0.7882}% #5294c9

\DeclareMathOperator{\Tr}{Tr}

\renewcommand{\vec}[1]{\boldsymbol{#1}}
\usepackage{youngtab}
\usepackage{slashed}

\definecolor{rossoCP3}{cmyk}{0,.88,.77,.40}

\begin{document}
%%%%%%%%%%%%%%%%%%%%%%%%%%%%%%%%%%%%%%%%%%%%%%%%%%%%%%%%%%%%%%%%%%%%%%%%%%%
\title{\Large  Schr\"odinger functional boundary conditions and
  improvement for $N>3$\color{rossoCP3}   }
\author{Ari Hietanen$^{\color{rossoCP3}{\varheartsuit}}$}\email{hietanen@cp3-origins.net} 
\author{Tuomas Karavirta$^{\color{rossoCP3}{\varheartsuit}}$}\email{karavirta@cp3-origins.net} 
\author{Pol Vilaseca$^{\color{rossoCP3}{\spadesuit}}$}\email{pol.vilaseca.mainar@roma1.infn.it} 

\affiliation{
\vspace{5mm}
{$^{\color{rossoCP3}{\varheartsuit}}${ \color{rossoCP3}  \rm CP}$^{\color{rossoCP3} \bf 3}${\color{rossoCP3}\rm-Origins} \& the {\color{rossoCP3} \rm Danish IAS},
University of Southern Denmark, Campusvej 55, DK-5230 Odense M, Denmark}
\vspace{5mm} \\
{$^{\color{rossoCP3}{\spadesuit}}${ \color{rossoCP3}  \rm Instituto Nazionale di Fisica Nucleare} 
Sezione di Roma,
P.le A. Moro 2, I-00185 Roma, Italia}
\vspace{1cm}
}

%%%%%%%%%%%%%%%%%%%%%%%%%%%%%%%%%%%%%%%%%%%%%%%%%%%%%%%%%%%%%%%%%%%%%%%%%%%%%%%%%%%%%%%%
\begin{abstract}
The standard method to calculate non-perturbatively 
the evolution of the running coupling of a SU($N$) gauge theory
is based on the Schr\"odinger functional (SF). In
this paper we construct a family of boundary fields for general values of $N$
which enter the standard definition of the SF coupling. 
We provide spatial boundary conditions for fermions in several representations
which reduce the condition number of the squared Dirac operator.
In addition, we calculate the $\mathcal{O}(a)$ improvement coefficients 
for $N>3$ needed to remove boundary cutoff effects from the gauge action. 
After this, residual cutoff effects on the step scaling function are shown
to be very small even when considering non-fundamental representations.
We also calculate the ratio of $\Lambda$ parameters 
between the $\overline{\textrm{MS}}$ and
SF schemes. 
\\ ~ \\
[.1cm]
{\footnotesize  \it Preprint: CP$^3$-2014-031 DNRF90 \& DIAS-2014-31}
\end{abstract}

\maketitle

\section{Introduction}

Asymptotically free theories, such as gauge theories coupled to fermionic
matter fields\footnote{For a small enough number of fermionic degrees of
  freedom.}, are characterized by having a coupling which becomes
small at short distances. This property enables reliable perturbative
calculations of physical quantities at large energies. A dimensionful scale is dynamically generated through  the process of
dimensional transmutation. Typically, this scale is associated in
perturbation theory with the $\Lambda$ parameter, i.e. a
multiplicative constant of the integrated beta function.

The non-perturbative evolution of the running coupling in different gauge field theories from the low energy 
sector to the high energy regime has been the central goal of many studies. The standard approach is the use of a finite size scaling
technique based on the Schr\"odinger functional (SF), in which the size
of the system is associated to the renormalization scale \cite{Luscher:1992an}. This method was 
successfully used to calculate the scale evolution of the coupling in the 
SU(2) \cite{Luscher:1992zx} and SU(3) \cite{Luscher:1993gh}
Yang-Mills theories and in QCD \cite{DellaMorte:2004bc}. Motivated 
by ideas of physics beyond the standard model (BSM), in the last decade this 
method has also been applied to study the SU(4) pure gauge theory \cite{Lucini:2008vi} and
several theories containing matter transforming under higher dimensional
representations of the gauge group or a large number of fermions in the fundamental representation
\cite{Appelquist:2007hu,Hietanen:2009az,Bursa:2009we,Bursa:2010xn,DeGrand:2010na,Karavirta:2011zg,DeGrand:2012qa,Hayakawa:2010yn,Heller:1997vh}. 

However, for lattices accessible in typical numerical simulations SF schemes
are affected by lattice artifacts arising from the bulk and from the 
boundaries of the lattice. These can be removed, following Symanzik's improvement
program, by adding the corresponding counterterms to the action at the 
bulk and the boundaries. Symanzik's program was successfully carried
out in
\cite{Luscher:1992an,Luscher:1996vw,Luscher:1996sc,Sint:1995ch,Sheikholeslami:1985ij},
where the improvement coefficients necessary to remove $O(a)$ effects from the
coupling were calculated in perturbation theory. 

For theories beyond QCD the situation is still inconclusive. A program for the
non-perturbative study of SU($N$) gauge theories in the large $N$ limit \cite{Lucini:2012gg} started
in the last decade driven by interest from string theory. As part of
that program, in \cite{Lucini:2008vi} the ratio $\Lambda_{\overline{\textrm{MS}}}/\sqrt{\sigma}$
between the lambda parameter in the $\overline{\textrm{MS}}$ scheme and the 
string tension $\sigma$ was calculated for the SU(4) theory aiming
to obtain extrapolations of  
the $N$ dependence of the $\Lambda_{\overline{\textrm{MS}}}/\sqrt{\sigma}$ 
in the large $N$ limit. There, the dominant 
systematic errors are due to the lattice artifacts present by using an 
unimproved action.
For the case of theories with non-fundamental fermions, although the $O(a)$
improvement coefficients are known, the remaining higher order cutoff effects
have been reported to be very large if the standard setups, which work fine
for QCD, are naively exported.

In the last few years a new renormalized coupling based on the gradient
flow (GF) has been proposed for step scaling studies
\cite{Narayanan:2006rf,Luscher:2010iy,Luscher:2011bx,Ramos:2013gda,Fritzsch:2013je}. Compared 
to the original SF coupling based on a background field (see bellow), the gradient flow coupling
has the advantage that considerably smaller statistics are required for obtaining
a similar accuracy.

However, there are some situations where the original SF coupling is
superior compared to the gradient flow. First of all, it has been
observed that while the GF coupling works better at large physical
volumes, at small volumes the SF coupling fares better than the
gradient flow \cite{Ramos:2014}. Also, in the pure gauge theory,
relevant for the large $N$ limit, the generation of 
configurations is so fast compared to the measurement of the gradient
flow that the reduced accuracy can be overcome with increased 
statistics. In addition, in BSM lattice studies one is often
interested in the existence of a nontrivial infra-red fixed
point. The value of the coupling constant $g$ at the fixed point is a
renormalization scheme dependent quantity and it differs between
Schr\"odinger functional and gradient flow schemes. Therefore, it is
possible that in a specific scheme the coupling is too strong at the
fixed point or it is on the wrong side of a bulk phase
transition. This is true even if the fixed point is visible in other
schemes. The only study, we are aware of, that compares these two
methods with the same action found the gradient flow coupling to be
about twice the Schr\"odinger functional coupling
\cite{Rantaharju:2013bva}. Moreover, due to the property of continuum
reduction \cite{Narayanan:2003fc}, at large $N$ it is possible to do
simulations at small  lattice volumes where the SF coupling is known
to perform well. 

This work completes the Schr\"odiger functional framework to study the
phase diagram of strongly interacting gauge theories
\cite{Sannino:2004qp} with any $N$ or representation. In the paper we
generalize the boundary conditions for the  gauge fields in the SF to
obtain a family of schemes useful for arbitrary $N$ 
with a good signal to noise ratio in lattice simulations.
Moreover, the $O(a)$ improvement coefficients are obtained to one loop order
in perturbation theory. For this, we calculate the one loop running 
coupling in our family of SF schemes following closely the discussions
in \cite{Luscher:1992an,Sint:1995ch} and  adapting them to arbitrary
$N$. The values obtained  for the boundary improvement coefficients
are valid for any choice of Dirichlet boundary conditions at the temporal
boundaries. With this knowledge we relate the
$\Lambda$ parameters between our SF schemes and the more widely used
$\overline{\textrm{MS}}$ scheme. Another appealing property of the
present family of schemes is that, together with an appropriate choice
of spatial boundary conditions for the fermions, they lead to a setup
for which higher order cutoff effects due to fermions are very small
even for non-fundamental representations. Preliminary results of this
work have been published in \cite{Karavirta:2013qqa}.

The paper is organized as follows: in section~\ref{sec:sf} we
recall some concepts concerning the Schr\"odinger functional and
collect a set of formulas useful for the remaining discussion. In
section~\ref{sec:boundary} the generalized boundary
conditions are provided. The calculation of the improvement coefficients is 
presented in
section~\ref{sec:improvements}, where we also discuss the effect that the 
fermionic spatial boundary conditions have on the residual higher order
cutoff effects. The matching of the $\Lambda$ parameters to the $\overline{\textrm{MS}}$
scheme is done in section~\ref{sec:matching}. We conclude in section~\ref{sec:conclusions}.

\section{Schr\"odinger functional \label{sec:sf}}
In this section we briefly recall the ideas introduced in \cite{Luscher:1992an,Sint:1995ch,Sint:1995rb}
 and collect the expressions necessary for the subsequent
discussion. We refer the interested reader to the original articles for further detail.

The Schr\"odinger functional is the euclidean propagation amplitude 
between a field configuration at time $0$ and another field configuration
at time $T$, which has a path integral representation given by
\begin{equation}
 \mathcal{Z}[C,C']=\int\mathcal{D}[U,\bar\psi,\psi]e^{S[U,\overline\psi,\psi]},
\end{equation}
with Dirichlet boundary conditions specified for the gauge fields $U$ and 
the fermion fields $\psi$ and $\bar\psi$.

In the present work, we are interested in the $\mathcal{O}(a)$ improved Wilson action
\be
S[U,\bar\psi,\psi]=S_G[U]+S_F[U,\psi,\bar\psi].
\label{eq:action}
\ee
The pure gauge part is the standard SU($N$) Wilson gauge action
\begin{equation}
  S_G[U] = \frac{1}{g_0^2}\sum_{P}w(P)\rm{Tr}[1-U(P)],
  \label{eq:gaugeact}
\end{equation}
The spatial components of the gauge fields at the temporal boundaries ($t=0$ and $t=T$)
satisfy nonhomogeneous Dirichlet boundary conditions
\begin{equation}
  U_k(t=0,\vec{x})=W_k(\vec{x}),\quad U_k(t=T,\vec{x})=W_k'(\vec{x}),\quad k=1,2,3.
\label{eq:gauge_bound}
\end{equation}
The boundary gauge fields $W_k$ and $W_k'$ can be parametrized as 
\begin{equation}
W_k(\vec{x})=\exp(aC_k(\eta)), \quad  W'_k(\vec{x})=\exp(aC'_k(\eta)),
\label{eq:gauge_bound2}
\end{equation}
where $C_k(\eta)$ and $C'_k(\eta)$ are taken to be homogeneous, abelian
and spatially constant \cite{Luscher:1992an}, and they depend on a
dimensionless parameter $\eta$. A specific form for these boundary matrices is derived
in section~\ref{sec:boundary} for gauge group SU($N$) with arbitrary $N$. In the spatial
directions the gauge filds are taken to be periodic $U_{\mu}(t,{\vec{x}})=U_{\mu}(t,{\vec{x}}+L\hat k)$.

The weight $w(P)=1$ except for the spatial plaquettes at the
boundaries for which $w(P)=\frac12$. Due to the particular choice of boundary
conditions for the gauge fields, the spatial boundary plaquettes give only a constant
contribution to the action and can be ignored. It is well known that within SF schemes,
the mere presence of temporal boundaries 
constitutes an extra source of lattice artifacts. %$\mathcal{O}(a)$ errors to the action. 
Removal of
these effects has first been studied in
\cite{Luscher:1992an,Luscher:1996vw,Luscher:1996sc}, where it was shown
that the $\mathcal{O}(a)$ lattice artifacts coming from the boundaries can be canceled 
by tuning the weight $w(p)=c_t(g_0)$ for the temporal plaquettes attached to the 
boundaries, where $c_t$ is the coefficient of a dimension 4 counterterm localized at the 
boundaries \cite{Luscher:1992an}.

The fermionic part of Eq.~(\ref{eq:action}) is the standard Wilson fermion action with the
clover term
\begin{equation}
  S_F[U,\psi,\bar\psi] =a^4\sum_x\bar{\psi}(x)(D_{WD}+m_{0})\psi(x),
  % \sum_{\NF} 
  %\sum_{x,y} \bar\psi_{f,x} M_{xy} \psi_{f,y} 
    \label{sf_impr}
\end{equation}
where $D_{WD}$ is the improved Wilson-Dirac operator
\be
D_{WD}=\frac{1}{2}[\gamma_\mu(D^*_{\mu}+D_{\mu})-aD^*_{\mu}D_{\mu}]+c_{sw}\frac{i a}{4}\sigma_{\mu\nu} F_{\mu\nu}(x).\label{wilsondirac}
\ee
The operator $F_{\mu\nu}(x)$ is the symmetrized lattice field strength
tensor, $\sigma_{\mu\nu}=\frac i2\left[\gamma_\mu,\gamma_\nu\right]$ 
and the operators $D_{\mu}$ and $D^* _{\mu}$ are the covariant forward and
backward derivatives which are defined in Eq.~\eqref{eq:covder}. The
improvement coefficient $c_{sw}$ can be determined perturbatively
\cite{Wohlert:1987rf,Luscher:1996vw} and non-perturbatively
\cite{Luscher:1996ug,Karavirta:2011mv}. To the lowest order in
perturbation theory $c_{sw}=1$
\cite{Sheikholeslami:1985ij}. The removal of
$O(a)$ effects arising from the interplay between fermions and the SF boundaries requires the addition of another dimension
4 counterterm at the boundaries. Since this does not contribute to the 
observables studied further in this work at the present order in perturbation
theory, we ignore it from now on and refer the reader to the original 
literature \cite{Luscher:1996sc} for further details.

The fermionic fields satisfy the following boundary conditions
\begin{align}
  \left.P_+ \psi\right|_{t=0} & = \left.P_- \psi\right|_{t=T} = 0, \\
  \left.\bar{\psi}P_-\right|_{t=0} & = \left.\bar{\psi}P_+\right|_{t=T} = 0,
\end{align}
where $P_{\pm} = \frac12(1\pm\gamma_0)$. The boundary conditions in the spatial
directions are periodic up to a phase 
\cite{Sint:1995rb}:
\begin{equation} \label{eq:spatial_bc}
  \psi(x+L\hat{k})=e^{i\theta_k}\psi(x),\qquad
  \bar\psi(x+L\hat{k})=\bar\psi(x)e^{-i\theta_k}.  
\end{equation}
The phase is usually chosen so that the smallest eigenvalue of the squared Dirac
operator is large \cite{Sint:1995rb}. In this situation, the condition number 
(i.e. the ratio between the highest an lowest eigenvalues) is small, which improves the speed
of the known inversion algorithms.
However, the value of $\theta$ also has an effect on the
convergence of the 1-loop perturbative
coupling to its continuum limit. This is discussed in subsection
\ref{sec:cutoff}.   

The boundary conditions for the gauge fields in Eqs.~(\ref{eq:gauge_bound}) and
(\ref{eq:gauge_bound2})
induce a constant chromo-electric background field $V_\mu(x)$ in the space-time. 
The variable $\eta$ in the boundary fields Eq.~(\ref{eq:gauge_bound2})
parametrizes a curve of background fields.
A renormalized coupling can be defined \cite{Luscher:1992an} as a response of the system 
to a deformation of the background field
\be
\left. \frac{\partial\Gamma}{\partial\eta} \right\vert_{\eta=0} = \frac{\kappa}{\overline{g}^2},
\label{eq:ren_coupling}
\ee
with the effective action $\Gamma =-\ln \mathcal{Z}$. The normalization 
constant 
\be
\kappa=\left.\frac{\partial\Gamma_{0}}{\partial\eta}\right|_{\eta=0},
\label{kappa}
\ee
is defined so that $\overline{g}^2=g_0^2$ to the lowest order of perturbation theory.

One of the central quantities in numerical simulations is the step scaling function
\begin{equation}
 \sigma(u)=\left.\overline{g}^{2}(2L)\right|_{u=\overline{g}^{2}(L)}.
\end{equation}
This is required for reconstructing non-perturbatively the scale evolution of the 
running coupling. In presence of a lattice regulator, the deviations of the 
lattice counterpart of the step scaling function
$\Sigma(u,L/a)$ from the continuum $\sigma(u)$ can be used to monitor the size of cutoff 
effects (see subsection \ref{sec:cutoff}).
%\begin{equation}
%  W_k(\vec{x})=\exp(aC_k(\eta)) \;\; \textrm{and} \;\;  W'_k(\vec{x})=\exp(aC'_k(\eta)),
%\end{equation}
%where $C_k(\eta)$ and $C'_k(\eta)$ are Abelian traceless
%anti-Hermitean matrices, 
%introduce a constant chromo-electric
%background field $V_\mu(x)$ to the space-time. The variable $\eta$
%parameterizes a curve of background fields. Given an effective action
%$\Gamma =-\ln \mathcal{Z}$ we can define a renormalized coupling as
%\be
%\left. \frac{\partial\Gamma}{\partial\eta} \right\vert_{\eta=0} = \frac{\kappa}{\overline{g}^2}.
%\label{eq:ren_coupling}
%\ee
%The 

\subsection{1-loop expansion}
The renormalized coupling Eq.~(\ref{eq:ren_coupling}) is suitable for both 
perturbative and non-perturbative evaluation. The 1-loop calculation 
of Eq.~(\ref{eq:ren_coupling}) was done in \cite{Luscher:1992an,Luscher:1993gh} 
for the pure
gauge theory in SU(2), and extended to accommodate fermions in \cite{Sint:1995rb}
\footnote{The evaluation of Eq.~(\ref{eq:ren_coupling}) to 2 loops was 
done in \cite{Bode:1998hd,Bode:1999sm}.}.
Non fundamental fermions have been considered in \cite{Karavirta:2012qd,Sint:2011gv,Sint:2012ae}.
In the present work we extend the previous calculations to arbitrary $N$. Although
the main strategy of the calculation follows closely previous works, some 
care has to be taken to generalize those ideas without complicating
the calculation. In the present subsection we collect some  
formulas necessary for the subsequent discussion and leave all technical details
on the calculation to Appendices~\ref{app:pertdetails} and \ref{app:basis}.

%The calculation of the perturbative 1-loop coupling follows
%\cite{Luscher:1992an} with certain modifications. Gauge fixing
%procedure adds two new terms to the action. These terms in the 1-loop
%level in perturbation theory are  
%\be
%S_{GF}[q]&=&\lambda_0 a^4\sum_{x,\mu,\nu} \Tr\left[ q_{\mu}(x)D _{\mu}D^* _{\nu}q_{\nu}(x) \right]+\mathcal{O}(g_0),\\
%S_{FP}[c,\bar{c}]&=&2 a^4\sum_{x,\mu} \Tr\left[ \bar{c}(x)D^* _{\mu}D_{\mu}c(x) \right]+\mathcal{O}(g_0).
%\ee  
%The functions $c(x)$ are the ghost fields and $q_{\mu}(x)$ are
%perturbations in the gluonic field with respect to a static background
%field $V_\mu(x)$ defined as 
%\be
%U_\mu(x)=\exp\left[g_0 q_{\mu}(x)\right]V_\mu(x)=\left(1+g_0 q_{\mu}(x)+\mathcal{O}(g_0 ^2)\right)V_\mu(x).
%\ee

%The path integral representation of the Schr\"odinger functional is 
%\be
%\mathcal{Z}\left[C(\eta),C'(\eta)\right]=\int \mathcal{D}\psi\mathcal{D}\bar{\psi}\mathcal{D}q\mathcal{D}c\mathcal{D}\bar{c}\exp\left(-S[q,\psi,\bar\psi,c,\bar{c}]\right)
%\ee
%This can also be written as an effective action 
After going through the gauge fixing procedure \cite{Luscher:1992an},
the effective action is expanded to 1-loop as
\be
%\Gamma= -\ln\mathcal{Z}=g_0^{-2} \Gamma_0+\Gamma_1+\mathcal{O}(g_0 ^2).
\Gamma= g_0^{-2} \Gamma_0+\Gamma_1+\mathcal{O}(g_0 ^2).
\label{efaction1}
\ee
Here $\Gamma_{0}$ is the classical action.
The 1-loop term $\Gamma_1$ in the effective action can be written as
\be
\Gamma_1=-\ln\det\Delta_0+1/2\ln\det\Delta_1-1/2\ln\det\Delta_2,
\ee
where $\Delta_{0}$, $\Delta_{1}$ and $\Delta_{2}$ are the quadratic ghost, gluonic and fermionic operators respectively. 
The explicit forms of the operators $\Delta_i$ are 
given in Appendix~\ref{app:pertdetails}.

%In Eq.~\eqref{efaction1} we have also written the perturbative expansion of the effective action to 1-loop order. The effective action can be used to define the running coupling
%\be
%\overline{g}^2(L/a)=\frac{\partial \Gamma_0/ \partial\eta}{\partial \Gamma/ \partial\eta}.
%\ee
%No we have a perturbative expression for the running coupling
The renormalized 
coupling Eq.~(\ref{eq:ren_coupling}) is also expanded in perturbation theory
\begin{equation}
\overline{g}^2(L/a)=g_0 ^2+p_1(L/a) g_0 ^4+\mathcal{O}(g_0 ^6).
\label{gbarexpansion} 
\end{equation}
%\begin{eqnarray}
%\overline{g}^2(L/a)&=&\frac{\partial \Gamma_0/ \partial\eta}{\partial (g_0 ^{-2}\Gamma_0+\Gamma_1)/ \partial\eta},\nonumber\\
%&=&g_0 ^2+p_1(L/a) g_0 ^4+\mathcal{O}(g_0 ^6),
%\label{gbarexpansion}
%\end{eqnarray}
According to Eq.~(\ref{efaction1}), the 1-loop coefficient
\be
p_1(L/a)= -\frac{\partial \Gamma_1/ \partial\eta}{\partial \Gamma_0/ \partial\eta},
\label{p_1}
\ee
%The perturbative
%expansion of the coupling constant is 
%given by Eq.~\eqref{gbarexpansion}. The 1-loop coefficient $p_{1}(L/a)$ 
receives an independent contribution from ghost, gauge, and fermionic fields
\begin{equation}
 p_{1}(L/a)=h_{0}(L/a)-\frac{1}{2}h_{1}(L/a)+\frac{1}{2}h_{2}(L/a)=p_{1,0}(L/a)+N_f p_{1,1}(L/a),
\label{eq:1loop_coupling}
\end{equation}
with
\be
 h_{s}&=&\frac{1}{\kappa}\frac{\partial}{\partial\eta}\textrm{ln} (\textrm{det}\Delta_{s}), \qquad s=0,1,2.
\label{eq:hs}
\ee

The gauge and fermionic contributions to Eq.~(\ref{eq:1loop_coupling}) 
can be calculated independently.
The gauge part is given by
\be
p_{1,0}(L/a)=h_{0}(L/a)-\frac{1}{2}h_{1}(L/a).
\label{eq:p10}
\ee
The calculation of $h_0(L/a)$ and $h_1(L/a)$ has been described in
great detail for SU(2) in \cite{Luscher:1992an} and the calculation
has been done for $N=3$ in \cite{Luscher:1993gh}. In
Appendix~\ref{app:pertdetails} we give the generalization of 
the calculations to $N\geq 3$. %and to an arbitrary choice of the
%background field.

The calculation of the fermionic  part $p_{1,1}(L/a)$ is 
straight forward to generalize to any boundary
fields and to any representation of the gauge group. One just needs
to replace the link variables in the Wilson Dirac operator Eq.~(\ref{wilsondirac})
with their counterparts in the desired representation.
Thus we will refer the interested reader to the original paper
\cite{Sint:1995ch}.

 The continuum and lattice step scaling functions are given 
to first order in perturbation theory by
\begin{equation}
 \sigma(u)=u+\sigma_{1}u^{2}+O(u^{3}),\qquad \Sigma(L/a,u)=u+\Sigma_{1}(L/a)u^{2}+O(u^{3}),
\end{equation}
with $\sigma_{1}=2b_{0}\ln(2)$.
The 1-loop coefficient $b_{0}$ of the beta function 
is given in an arbitrary representation by
\begin{equation}
 b_{0} = b_{0,0}+N_{f}b_{0,1},\qquad b_{0,0}=\frac{1}{(4\pi)^{2}}\frac{11}{3}C_2(F), \qquad b_{0,1}=-\frac{1}{(4\pi)^{2}}\frac{4}{3}T_{R},
\end{equation}
where the color group invariants are defined as
\be
C_2(A)\delta^{AB}=f^{ACD}f^{BCD},\qquad T_R\delta^{AB}=\Tr[t^A t^B],\qquad C_2(R)=t^{A}t^{A},
\ee
in the representation $R$ of SU($N$)~\footnote{The
values of the invariants are given by $T_{F}=1/2$, $C_2(F)=N$, $T_{A}=N$, $C_2(A)=(N^2-1)/(2 N)$, $T_{S}=(N+2)/2$, $C_2(S)=(N-1)(N+2)/N$, 
$T_{AS}=(N-2)/2$ and $C_2(AS)=(N+1)(N-2)/N$ for the fundamental, adjoint, symmetric and antisymmetric 
representations respectively.}.

Similarly as in Eq.~(\ref{eq:1loop_coupling}), the step scaling functions can be separated
into a gauge and a fermionic part,
\begin{equation}
 \sigma_{1}=\sigma_{1,0}+N_{f}\sigma_{1,1},\qquad \Sigma_{1}(L/a)=\Sigma_{1,0}(L/a)+N_{f}\Sigma_{1,1}(L/a).
\end{equation}
This allows us to study separately cutoff effects due to gauge and fermion fields independently.

%We calculate $p_1$ as a function of the lattice size $L/a$ by
%evaluating the determinant of the operators $\Delta_s$ by solving
%their eigenvalues.
% This is equivalent to solving a difference relation involving the operators $\Delta_s$. 
%In this way we determine  $p_1(L/a)$ and consequently the running coupling $\overline{g}^2$ to one loop order in
%perturbation theory for a range in $L/a$.

\section{Boundary Fields for $N>2$}\label{sec:boundary}
In this section we present a generalization of the
boundary fields for $N>2$. The selection of the boundary fields is
only limited by the requirement that there is a unique and stable
classical solution to system. In practice, this limits us to Abelian
boundary fields $W_k$ and $W_k'$ which can be written as in Eq.~\eqref{eq:gauge_bound2}, where
\begin{equation}
  C_k=\frac iL
  \begin{pmatrix}
    \phi_1 & 0 & \dots & 0\\
    0 & \phi_2 &  \dots & 0\\
    \vdots &\vdots & \ddots & \vdots \\ 
    0 & 0 & \dots &  \phi_N 
  \end{pmatrix}
 \;\; \textrm{and} \;\;
 C'_k = \frac iL
 \begin{pmatrix}
    \phi'_1 & 0 & \dots & 0\\
    0 & \phi'_2 &  \dots & 0\\
    \vdots &\vdots & \ddots & \vdots \\ 
    0 & 0 & \dots &  \phi'_N 
 \end{pmatrix}.
\end{equation}
Since $W_k$ has to be an SU($N$)-matrix the vectors  $\vec{\phi} =
(\phi_1,\phi_2,\dots,\phi_N)$ and $\vec{\phi}' =
(\phi'_1,\phi'_2,\dots,\phi'_N)$ must satisfy
\begin{equation}
 \sum_{k=1}^{N} \phi_k = 0 .
 \label{eq:constsum}
\end{equation}
Now the classical solution, i.e. the background field, can be written as
\begin{equation}
  V_\mu(x) = \exp(a B_\mu(x)),
  \label{background field}
\end{equation}
where
\begin{align}
  B_0(x) & = 0, \label{Bdefinition1}\\
  B_k(x) & = [x^0 C'_k+(L-x^0)C_k]/L.
  \label{Bdefinition2}
\end{align}
It is shown in \cite{Luscher:1992an} that the solution $V_\mu(x)$ is
absolutely stable if the vectors $\vec{\phi}$ and $\vec{\phi}'$
satisfy Eq.~\eqref{eq:constsum} and
\begin{align}
 \phi_1 < \phi_2<\dots&<\phi_N ,\\
  \phi_N-\phi_1 &< 2\pi.
  \label{eq:fundamentadomain}
\end{align}
These conditions define a fundamental domain, which is an irregular
$(N-1)$-simplex and has vertices at points
\be
\begin{array}{lcl}
  \vec{X}_1 &= &\frac{2\pi}{N}\left(-N+1,1,1,\dots,1\right),\\
  \vec{X}_2 & = &\frac{2\pi}{N}\left(-N+2,-N+2,2,\dots,2\right), \\
  \vec{X}_3 & = &\frac{2\pi}{N}\left(-N+3,-N+3,-N+3,3,\dots,3,\right), \\
 &  \vdots  & \\
  \vec{X}_{N-1} & = & \frac{2\pi}{N}\left(-1,-1,\dots,-1,N-1\right),\\
  \vec{X}_N & = & (0,0,\dots,0).
\end{array}
  \label{eq:nsimplex}
\ee

To define a renormalized coupling we can choose any two different points
inside the fundamental domain to set up the boundary fields. A
different choice leads to a different renormalization scheme, which
can be matched to each other using perturbation theory (see
section~\ref{sec:matching}). However, there are practical
considerations in selecting the boundary fields, namely the signal to
noise ratio in the Monte Carlo simulations and the size of higher
order lattice artifacts. Our choice is based on the attempt to
maximize the signal to noise ratio as in practice the minimization of
the higher order lattice artifacts often leads to a low signal, which
neglects the gains of a better continuum extrapolation.

To obtain a maximal signal strength we have two competing
requirements. We need to twist the gauge fields as much
as possible while  staying away from
the boundaries of the fundamental domain. This is because the coupling
is proportional to the twist and because closeness of the  instability
of the classical solution  increases noise. According to these considerations
we choose $\phi$ to be in the middle of a line connecting $\vec{X}_1$
and the centeroid of the fundamental domain
\begin{align}
  \vec{\phi} & =  \frac12\vec{X_1}+\frac{1}{2N}\sum_{k=1}^{N}\vec{X}_k\nonumber\\
  & = \frac{\pi}{2N}\left(3-3N,5-N,7-N,\dots,N+1\right).
\end{align}

To determine $\vec{\phi}'$ we find a transformation which is a map from the
fundamental domain to itself and mirrors the vertices. First we define
a simple map $R_{i,j}(\vec{\phi})$  that reflects the points in the fundamental domain
with respect to a $(N-2)$ dimensional hyperplane. The hyperplane
$R_{i,j}(\vec{\phi})$ goes through vertices $\vec{X}_k$, $k\neq i,j$  and
intersects the line connecting $\vec{X}_i$ and $\vec{X}_j$ at the
middle. For $N>3$ the function $R_{i,j}(\vec{\phi})$ is not in general a
mapping from the fundamental  domain to itself, but we can define a
composite mapping 
\begin{equation}
\mathcal{M}(\vec{\phi})=\left(R_{1,N-1}\circ R_{2,N-2}\circ\ldots\circ
R_{[N/2],N-[N/2]}\right)(\vec{\phi}),
\label{M}
\end{equation}
where $R_{i,i}(\vec{\phi})$ is the identity mapping and  $[x]$ denotes the
integer part of $x$. Now $\mathcal{M}(\vec{\phi})$ is  a mapping from the fundamental 
domain to itself and written in components it has a simple form 
\begin{equation}
  \phi'_i= \left[\mathcal{M}(\vec{\phi})\right]_i = -\phi_{N-i+1}.
\end{equation}

To define the coupling we choose a one parameter curve of background
fields $\vec{\phi}+\vec{t}(\eta)$. We select it in a way that the
results are equivalent to those of the SU(3) theory given in
\cite{Luscher:1993gh}\footnote{Note that the  boundaries are
  trivially   rotated compared to the ones in \cite{Luscher:1993gh}.}, i.e. we select $\vec{t}(\eta)$ so that
it changes sign under the mapping $\mathcal{M}(\phi)$ and points towards the boundaries of
the fundamental domain
\begin{align}
  \vec{t}(\eta) & = \frac{\eta N}{2\pi(N-2)} \left(\vec{X}_1-\vec{X}_{N-1}\right)\nonumber\\
  & = \left(-\eta,\frac{2\eta}{N-2},\dots,\frac{2\eta}{N-2},-\eta\right).
\end{align}

\begin{figure}[h]
\centering
\includegraphics[scale=0.77,trim={6cm 12cm 2cm 3cm},clip]{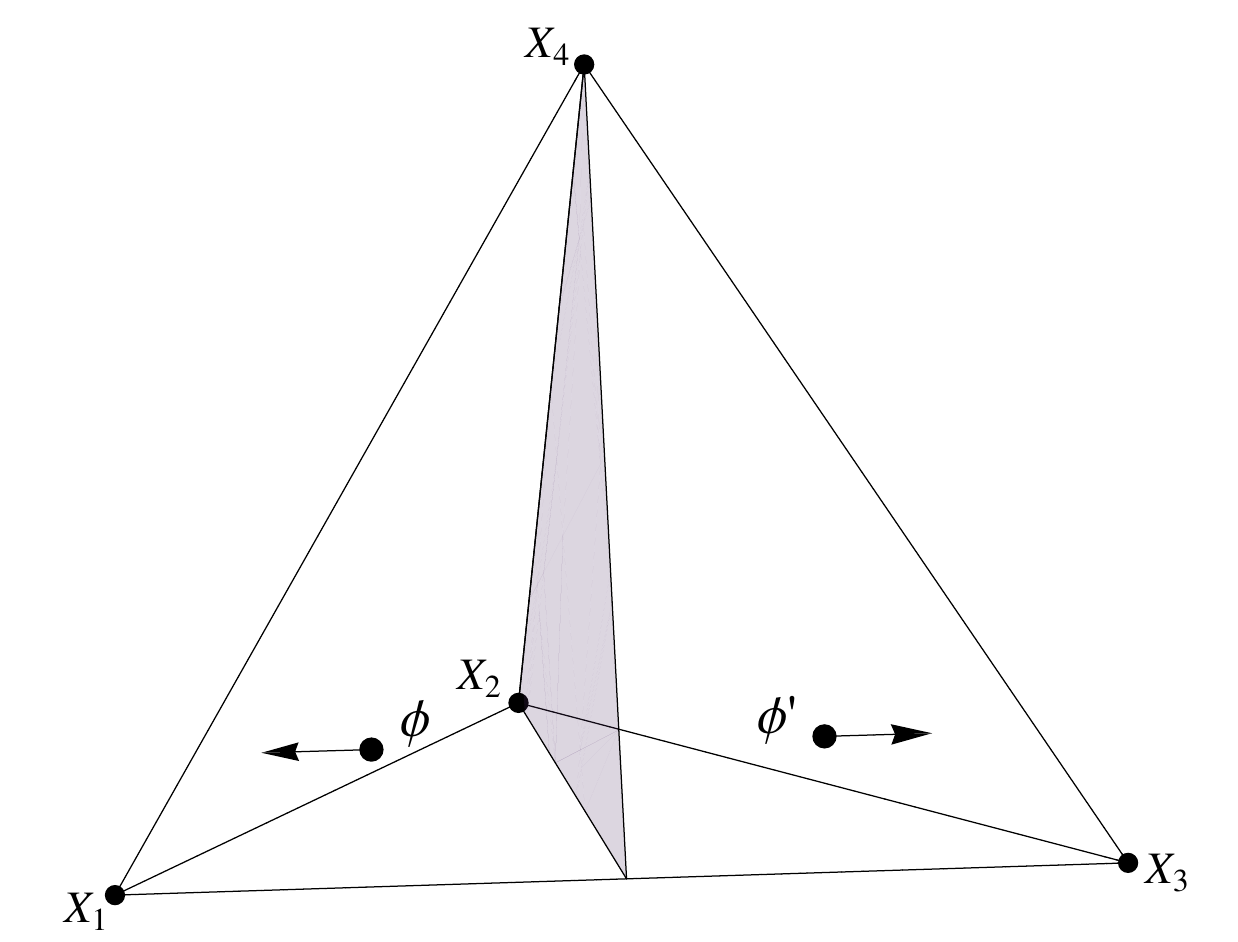}
\caption{Fundamental domain of SU(4). \label{fig:hyperplane}}
\end{figure}

See Fig.~\ref{fig:hyperplane} for illustration of the fundamental domain and boundary
conditions for SU(4) and Table~\ref{tab:boundaries} for the boundary
values for $N=3,4,5$.

In the lattice computations it is advisable {\emph{not}} to set $\vec{t}(\eta)$
beforehand, but to measure a complete $N-1$ dimensional basis which
can be used to construct a generic curve. Each curve corresponds to a
different renormalization scheme.

% with a specific choice of a
%curve the statistical errors may be reduced. 

\begin{table}
\begin{tabular}{ccc}
\hline
\multicolumn{3}{c}{$N=3$} \\
\hline
 $\phi$ && $\phi'$\\
\hline\hline
 $-\eta - \pi$  & \;\;\; & $\eta - \frac{2\pi}{3}$\\
 $2\eta + \frac\pi3$ & &  $-2\eta - \frac\pi3$\\
 $-\eta + \frac{2\pi}{3}$ & &  $\eta + \pi$\\
\hline
\\
\\
\end{tabular}
\;\;\;\;\;\;
\begin{tabular}{ccc}
\hline
\multicolumn{3}{c}{$N=4$} \\
\hline
$\phi$ && $\phi'$\\
\hline\hline
$ -\eta - \frac{9}{8}\pi$ & \;\;\;&$\eta - \frac{5}{8}\pi$\\
$  \eta + \frac{1}{8}\pi$ &       & $-\eta - \frac{3}{8}\pi$\\
$  \eta + \frac{3}{8}\pi$ &       & $-\eta - \frac{1}{8}\pi$\\
$ -\eta + \frac{5}{8}\pi$ &       & $\eta +  \frac{9}{8}\pi$\\
\hline
\\
\end{tabular}
\;\;\;\;\;\;
\begin{tabular}{ccc}
\hline
\multicolumn{3}{c}{$N=5$} \\
\hline
$\phi$ && $\phi'$\\
\hline\hline
$-\eta - \frac{6}{5}\pi$        &\;\;\; &  $\eta - \frac{3}{5}\pi$\\
$ \frac23\eta$                  & &  $-\frac23\eta - \frac{2}{5}\pi$\\
$\frac23\eta + \frac{1}{5}\pi$  & & $-\frac23\eta - \frac{1}{5}\pi$\\
$\frac23 \eta + \frac{2}{5}\pi$ & &  $-\frac23\eta$ \\
$-\eta + \frac{3}{5}\pi$        & &  $\eta +\frac{6}{5}\pi$\\
\hline
\end{tabular}
\caption{The values of the boundary fields for $N=3,4,5$.\label{tab:boundaries}}
\end{table}

\section{Boundary effects and improvements}\label{sec:improvements}

\subsection{Fermionic spatial boundary conditions}\label{subsec:theta}

Recalling the spatial boundary conditions for the fermion fields Eq.~(\ref{eq:spatial_bc}), 
we still have to choose a particular value for the angles
$\theta_{k}$. For simplicity, we consider the same angle in all
spatial directions $\theta=\theta_{k}$, $k=1,2,3$. 

We then fix $\theta$, following the criteria introduced in \cite{Sint:1995ch}, so that the minimum
eigenvalue $\lambda_{\textrm{min}}$ of the fermion operator $\Delta_{2}$ is as large as possible. 
This leads to a small condition number which optimizes the speed of the numerical inversion
of the operator. 

\begin{figure}
  \begin{center}
  \includegraphics[width=0.9\textwidth]{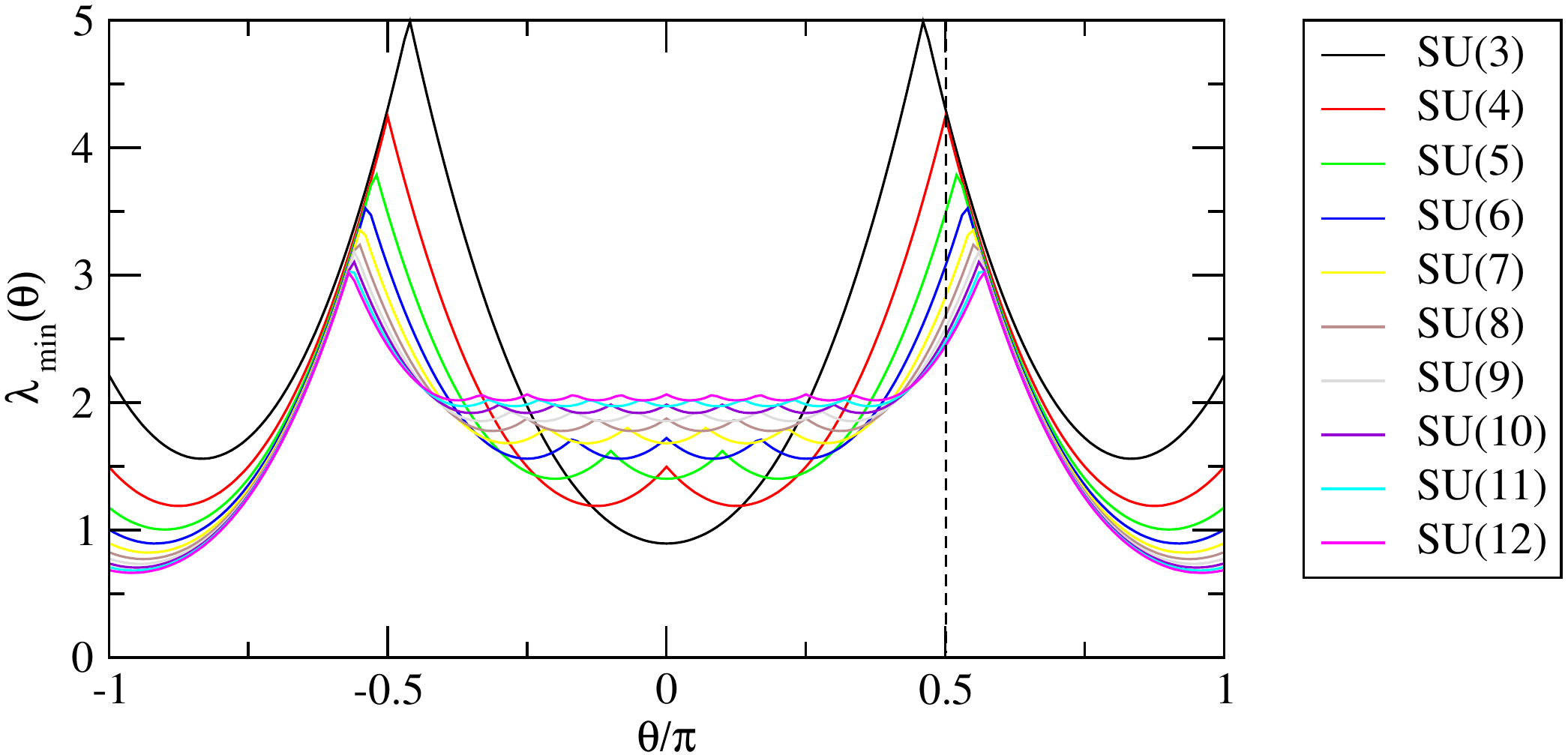}
  \caption{Lowest eigenvalue $\lambda_{\textrm{min}}$ (in units of $L^{-2}$ ) as a 
function of $\theta$ for the fundamental
representation of SU($N$), with $N\subset[3,12]$. The vertical discontinuous line marks the chosen value $\theta=\pi/2$.
\label{fig:lambda_theta}}
  \end{center}
\end{figure}

The values of $\theta$ leading to a maximum $\lambda_{\textrm{min}}$
depend on the background field and also on the fermion representation being 
considered. For the fundamental representation, the profile of smallest 
eigenvalues $\lambda_{\textrm{min}}$ as a function of $\theta$ is shown in
Fig.~\ref{fig:lambda_theta}  for the different gauge groups
considered in this work excluding the case of SU(2). 

Although the maximum of $\lambda_{\textrm{min}}$ is achieved at different values of $\theta$ for
every gauge group considered, the choice $\theta=\pi/2$ is always close to the maximum
and hence leads to a small condition number. For homogeneity in the
definition of a renormalization 
scheme in the subsequent calculations, we will fix $\theta=\pi/2$ for all 
values of $N$. As we will show
in Section~\ref{sec:cutoff}, this choice of $\theta$ together with the family of background
fields defined in this work will lead to a setup for which higher order cutoff effects are
highly suppressed even for non-fundamental representations.
Although the choice of $\theta=\pi/2$ is taken considering the fundamental 
representation, this value also leads to reasonably small condition numbers
for the symmetric and antisymmetric representations. 
For the adjoint representation the 
smallest condition number is obtained for $\theta=0$. However, we decide to stick to the 
choice $\theta=\pi/2$ also in this case since it leads to a situation were higher order 
lattice artifacts are highly reduced\footnote{We consider the reduction of
higher order cutoff effects, which have been shown to be very large \cite{Karavirta:2012qd,Sint:2011gv,Sint:2012ae}, to be of higher importance.}.

For the case of SU(2), we choose $\theta=0$ for the fundamental representation but leave 
$\theta=\pi/2$ for the symmetric/adjoint.

\subsection{Gauge boundary improvement}\label{subsec:gauge_impro}

\subsubsection{Expansion}

The variables $p_{1,i}(L/a)$ are expected to have an 
asymptotic expansion in $L/a$ \cite{Luscher:1992an}
\be
p_{1,i}(L/a)\sim\sum_{n=0} ^{\infty} (r_{n,i} + s_{n,i} \ln(L/a))\left(\frac{a}{L}\right)^{n},
\label{eq:p1i_series}
\ee
where $s_{0,i}=2 b_{0,i}$ and $s_{1,i}=0$ after setting $c_{\textrm{SW}}$
to its tree level value. The boundary improvement 
coefficients $c_t ^{(1,i)}$ 
are determined by demanding linear cutoff effects to 
be absent in Eq.~\eqref{eq:p1i_series}, which is achieved by fixing
$c_t ^{(1,i)} = r_{1,i}/2$. The continuum coefficients $r_{0,i}$
are needed when matching the $\Lambda$ parameter to other schemes. 

In order to extract the coefficients $r_{n,i}$ as accurately as possible
we first evaluate $p_{1,0}(L/a)$ and $p_{1,1}(L/a)$ adapting the strategies
in \cite{Luscher:1992an,Sint:1995ch} to general $N$ (see appendices \ref{app:pertdetails} and \ref{app:basis} for details on 
the calculation). Once the series of data for $p_{1,i}(L/a)$ is produced, the 
coefficients $r_{n,i}$ can be extracted using a suitable fitting procedure.

In the pure gauge case, we calculated
$p_{1,0}(L/a)$ for
 values of $L/a\in
\{6,8,\ldots,100\}$  and then used the "Blocking" method described in
\cite{Luscher:1985wf} to obtain the values of the asymptotic coefficients. The
calculation was done using floating point precision with 50 decimal
places for $2\le N\le 8$ and with quadruple precision for $N>8$. To
control the error we compared the results and errors obtained with
different level of accuracy. 
Since the asymptotic form Eq.~\eqref{eq:p1i_series} is expected to be 
valid as $a/L\rightarrow0$, we consider only values of 
$L/a\subset[28,100]$ when extracting the coefficients $r_{0,0}$ and
$r_{0,1}$. This choice produced the most reliable values for the 
coefficients and their relative errors. As a check we also reproduced
the known value of $s_{0,0}=2 b_{1,0}$ to a similar degree of accuracy. 

%{\textcolor{red}{Did we reproduce correctly the values of s00 and s10 ????}

%We found out that the extracted coefficients and
%their errors did not agree if the full range of $L/a$ was used. This
%happens because the expansion in Eq.~\eqref{eq:p10series} is only
%valid in the limit $a/L\rightarrow0$. Keeping this in mind we used
%$L/a\in\{28,30,\ldots,100\}$ which produced the most reliable 
%values for the coefficients and their relative errors.   

Concerning the fermionic part, values for $p_{1,1}(L/a)$ were produced 
at quadruple precision in the range $L/a\subset[4,64]$ (for even and odd
values) for all gauge groups and representations considered in this work.
This was enough to obtain the asymptotic coefficients in Eq.~(\ref{eq:p1i_series})
to very high precision (see tables \ref{table:coeffs_1} and \ref{table:coeffs_2}).
\subsubsection{Results}\label{sec:results}
\begin{table}
\center
\begin{tabular}{ccc}
\hline
$N$ & $r_{0,0}$ & $r_{1,0}$   \\
\hline
\hline
2 &0.202349528(3) & -0.108735(17)  \\
3 &0.368282146(3) & -0.177987(14)  \\
4 &0.520970830(2) & -0.244261(14)  \\
5 &0.673474985(2) & -0.309345(13)  \\
6 &0.826895868(3) & -0.373834(13)  \\
7 &0.981591358(3) & -0.437984(13)  \\
8 &1.137655320(3) & -0.501921(12)  \\
9 &1.295080018(5) & -0.565696(18)  \\
10& 1.45381790(5) & -0.629390(18)  \\
11& 1.61380703(5) & -0.693011(14)  \\
12& 1.77498215(5) & -0.756579(14) \\
\hline
\end{tabular}
\caption{Values of the pure gauge coefficients $r_{0,0}$ and $r_{1,0}$ for $N=2,\ldots,12$.\label{tab:icoeff_gauge}}
\label{table:coeffs_1}
\end{table}

\begin{table}
\center
\begin{tabular}{ccccc}
\hline
$N$ & Fundamental & Adjoint & Symmetric & Antisymmetric  \\
\hline
\hline
2  & -0.00342666(1) & -0.13787329(4) & -0.13787329(4) & -\\
3  & -0.00343842(1) & -0.20761772(4) & -0.17327682(5) & -0.00343842(1)\\
4  & -0.00344138(2) & -0.27682313(6) & -0.20788611(5) & -0.06893677(2)\\
5  & -0.00344277(1) & -0.34620010(5) & -0.24257541(4) & -0.10364831(4)\\
6  & -0.00344355(1) & -0.41563817(4) & -0.27729937(5) & -0.13840671(4)\\
7  & -0.00344406(1) & -0.48510052(4) & -0.31204105(4) & -0.17317365(5)\\
8  & -0.00344441(1) & -0.55457355(5) & -0.34679040(5) & -0.20793990(4)\\
9  & -0.00344468(1) & -0.62405117(5) & -0.38154235(4) & -0.24270438(2)\\
10 & -0.00344489(1) & -0.69353094(5) & -0.41629290(6) & -0.27746570(5)\\
11 & -0.00344506(1) & -0.76295832(5) & -0.45104376(5) & -0.31122243(3)\\
12 & -0.00344520(1) & -0.83249201(5) & -0.48579212(5) & -0.34698044(4)\\
\hline
\end{tabular}
\caption{Values of the fermionic coefficient $r_{0,1}$ for the fundamental, adjoint, symmetric and antisymmetric representations
of $N=2,\ldots,12$.\label{tab:icoeff_fermion}}
\label{table:coeffs_2}
\end{table}

\begin{figure}
  \begin{center}
  \includegraphics[width=0.8\textwidth]{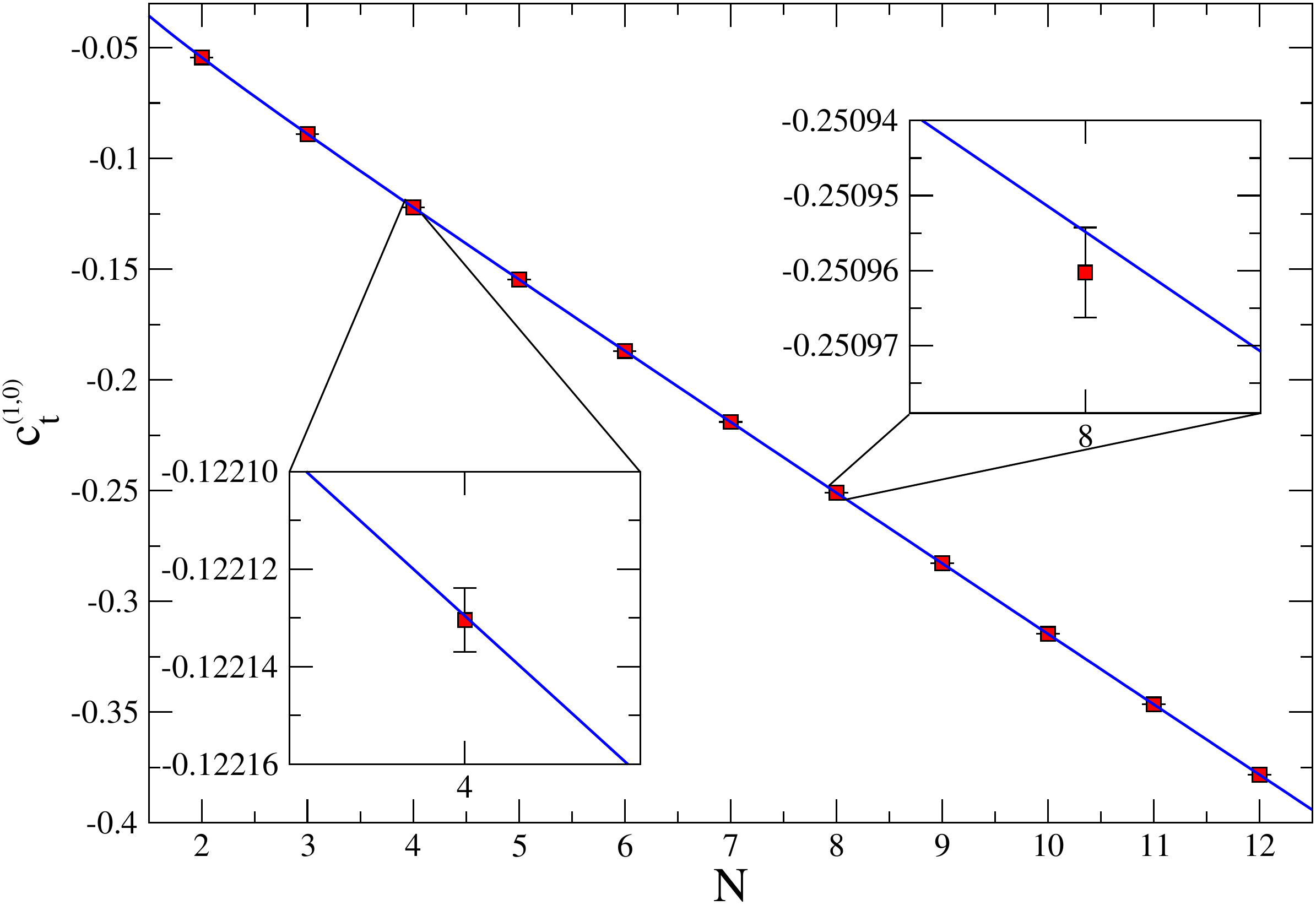}
  \caption{A polynomial fit to the $c_t^{(1,0)}$ data. We have zoomed out
    two points to illustrate the accuracy of the fit.\label{fig:ctfit}}
  \end{center}
\end{figure}

In Table~\ref{tab:icoeff_gauge} we give the values for the coefficients
$r_{0,0}$ and $r_{1,0}$. From $r_{1,0}$ we can extract the gauge contribution
to the boundary improvement
coefficient $c_t ^{(1,0)}=\frac12r_{1,0}$. According to continuum perturbation
theory we expect $c_t ^{(1,0)}$ to depend on group 
theoretical factors with the functional form 
\begin{equation}
  c_t ^{(1,0)} = A C_2(F) + B C_2(A) \equiv aN+\frac{b}{N},
  \label{eq:fitfun}
\end{equation}
where $C_{2}(R)$ is the quadratic Casimir operator in the
representation $R$. A fit to the data gives $b=0.017852(13)$ and $a=-0.0316483(4)$ with
an excellent $\chi^2/{\rm d.o.f.}\approx2.2/9\approx0.24$. The data and the fit are shown in  
Fig.~\ref{fig:ctfit}. To check the consistency of our results, we have
also performed fits adding additional terms to Eq.~(\ref{eq:fitfun}). 
The coefficients of the
additional terms are zero within statistical errors as shown in Table~\ref{tab:fits}.
 
For completeness, we also include here the value of the fermionic part 
$c_{t}^{(1,1)}$ for an arbitrary group and representation
\begin{equation}
 c_{t}^{(1,1)}(R)=0.038282(2)T_{R}
\end{equation}
 This was calculated for the fundamental representation in \cite{Sint:1995ch}
and later extended to other represesentations in \cite{Karavirta:2012qd}.
In the present work we have been able to reproduce the value of 
$c_{t}^{(1,1)}$ with similar accuracy, which is a further check on the 
correctness of the whole calculation.

\begin{table}
\begin{center}
  \begin{tabular}{crlc}
    \hline
    Fit function & \multicolumn{2}{c}{Parameters} & $\chi^2/\rm{d.o.f.}$\\
    \hline\hline
    \multirow{2}{*}{$a N + b/N$} & $a$ & $=-0.0316483(4)$ & \multirow{2}{*}{2.2/9=0.24} \\
    & $b$&$=0.017852(13)$ & \\
    \hline
    \multirow{3}{*}{$a N + b + c/N$}\quad & $\quad\quad a$ & $=-0.0316469(14)\quad\quad$ & \multirow{3}{*}{1.0/8=0.13} \\
    & $b$&$=-1.7(16)\times10^{-5}$  &\\
    & $c$&$=0.01789(4)$  &\\
    \hline
    \multirow{5}{*}{$a N^2 + bN + c + d/N+e/N^2$}\quad & $\quad\quad a$ & $=-2(135)\times10^{-8}\quad\quad$ & \multirow{5}{*}{0.76/6=0.13} \\
    & $b$&$=-0.03164(4)$  &\\
    & $c$&$=-5(26)\times10^{-5}$  &\\
    & $d$&$=0.0180(9)$  &\\
    & $e$&$=-2(8)\times10^{-5}$  &\\
    \hline
  \end{tabular}
\caption{Fits with a different functional forms for
  $c_t^{(1,0)}$.\label{tab:fits}}
\end{center}
\end{table}

\subsection{Residual cutoff effects\label{sec:cutoff}}

The determination of the gauge and fermion contributions to $c_{t}^{(1)}$ removes $O(a)$ lattice artifacts coming from the boundaries to
1-loop in perturbation theory. However, cutoff effects of higher order in $a$ are still present. We quantify these using the relative
deviations from the pure gauge and pure fermionic lattice step scaling
functions to one loop order, with respect to their universal 
continuum counterparts
\begin{equation}
 \delta_{1,i}(a,L)=\frac{\Sigma_{1,i}(L/a)-\sigma_{1,i}}{\sigma_{1,i}},\qquad \Sigma_{1,i}(L/a)=p_{1,i}(2L/a)-p_{1,i}(L/a).
\end{equation}

\begin{figure}
\centering
\includegraphics[width=0.8\textwidth]{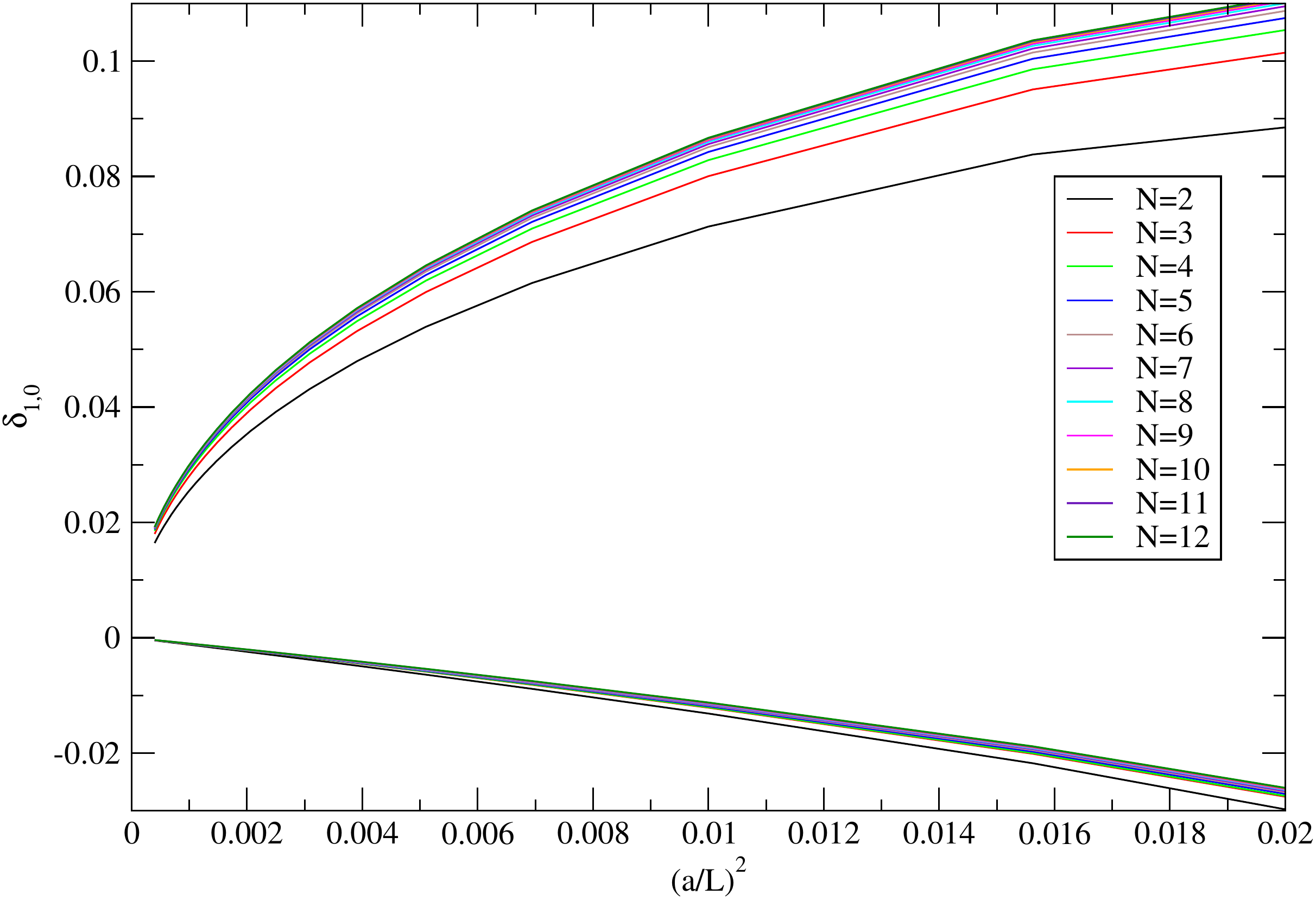}
\caption{Cutoff effects in the gauge part of the 1-loop step scaling function
   with $c_t ^{(1,0)}=0$ (upper set of lines) and with $c_t ^{(1,0)}$ set to its perturbative value (lower set of lines). \label{fig:cutoff_gauge}} 
\end{figure}

In Fig.~\ref{fig:cutoff_gauge} we show the convergence of the gauge
part of the 1-loop step scaling function with and without
improvement. It can be seen immediately that the residual cutoff
effects after improvement are quadratic in $(a/L)$ and small (order of
1\%) at $L/a=10$). Also the result depends only mildly on $N$.

\begin{figure}
  \begin{center}
  \includegraphics[width=1.0\textwidth]{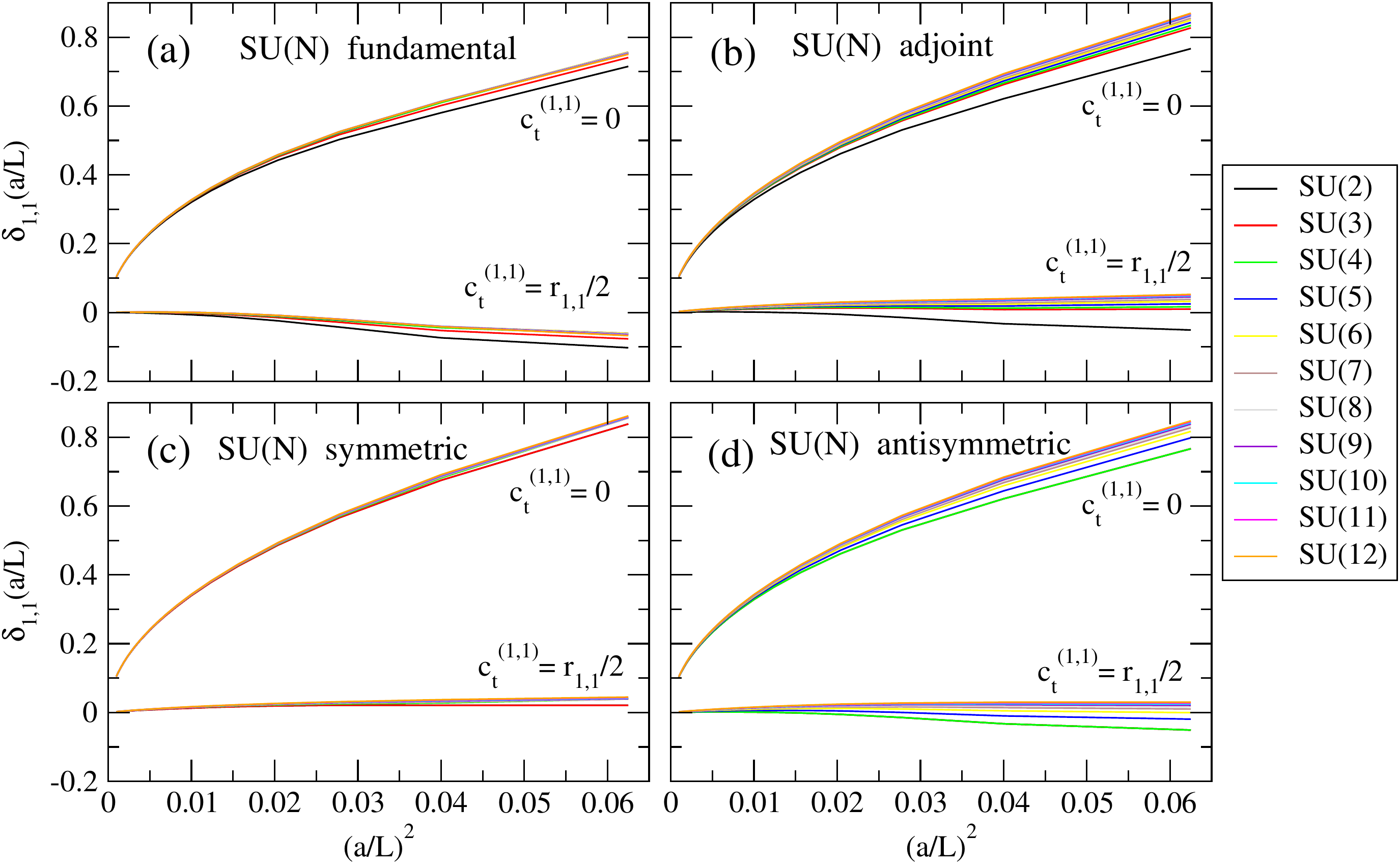}
  \caption{Cutoff effects in the fermionic part of the 1-loop step scaling
function due to a single flavor in the fundamental (a), adjoint (b),
symmetric (c) and antisymmetric (d) representations for the gauge groups
considered in this work. Cutoff effects are shown before ($c_{t}^{(1,1)}=0$)
and after ($c_{t}^{(1,1)}=r_{1,1}/2$) implementing $O(a)$ boundary
improvement. 
\label{fig:fermionic_cutoff_effects}}
  \end{center}
\end{figure}

The cutoff effects due to fundamental fermions are displayed in panel (a)
of Fig.~\ref{fig:fermionic_cutoff_effects}. From there, it is clear that 
in all the gauge groups considered in this work, the residual higher order cutoff effects 
are rather small after boundary $O(a)$ improvement is implemented. Residual cutoff effects
are of the order of $10\%$ already at the coarsest lattices considered, and converge to zero very 
fast. 

\begin{figure}
  \begin{center}
  \includegraphics[width=1.0\textwidth]{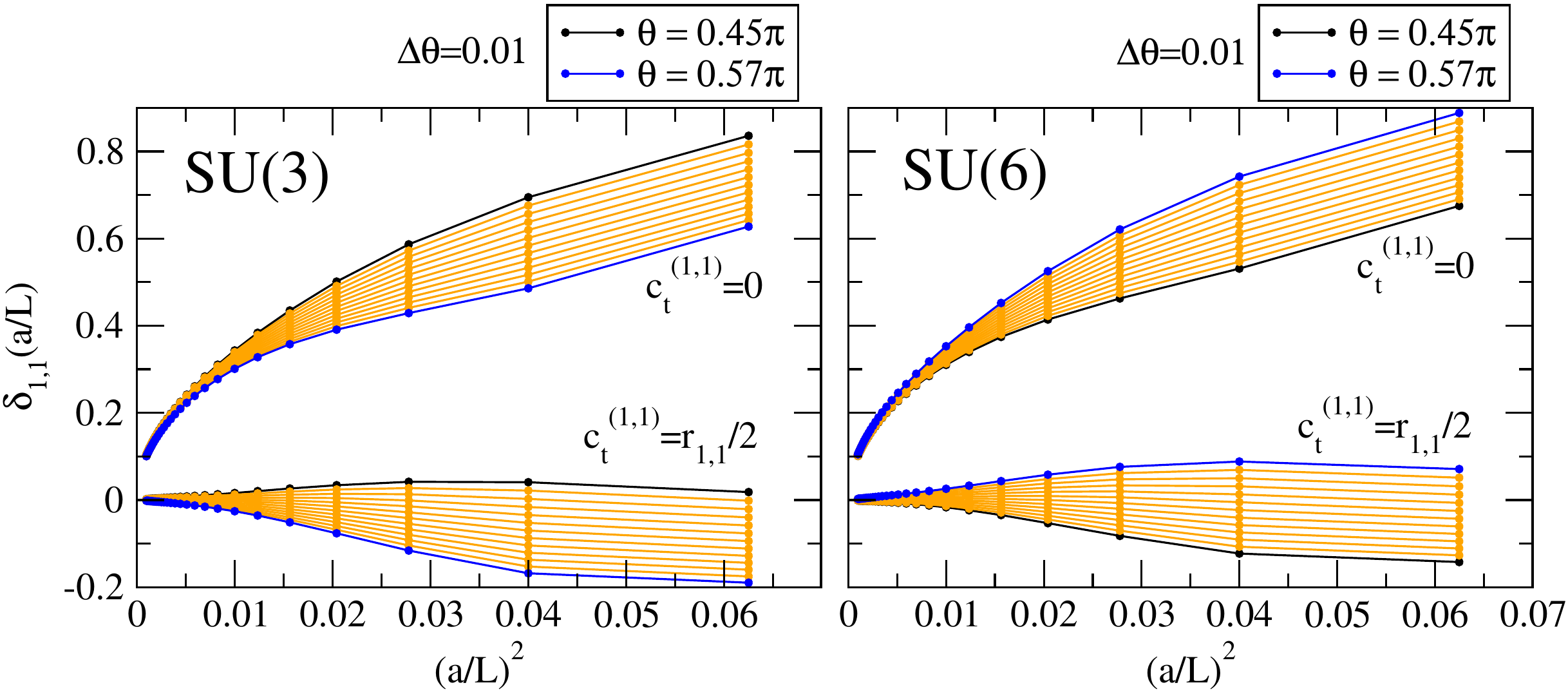}
  \caption{Cutoff effects in the fermionic part of the 1-loop step scaling
function for $N=3$ and 6. Cutoff effects are shown before ($c_{t}^{(1,1)}=0$)
and after ($c_{t}^{(1,1)}=r_{1,1}/2$) implementing $O(a)$ boundary
improvement.}
  \label{fig:cutoff_effects_theta_su3_su6}
  \end{center}
\end{figure}

This is true for the family of background fields defined in section \ref{sec:boundary} and for the value of $\theta$ chosen
in section \ref{subsec:theta}. A different choice of parameters, however, can lead to very high residual cutoff effects even after 
boundary $O(a)$ improvement is implemented\footnote{See references 
\cite{Karavirta:2012qd,Sint:2011gv,Sint:2012ae} for this issue in representations other than 
the fundamental.}. In order to check this, we study the dependence of $\delta_{1,1}$ on the parameter $\theta$ 
for different values of $N$ in a range $\theta\subset[0.45\pi,0.57\pi]$. The cases of $N=3$ and $6$ are displayed
in Fig.~\ref{fig:cutoff_effects_theta_su3_su6}. Other gauge groups show a very similar behavior. The residual cutoff effects $\delta_{1,1}$ depend
strongly on $\theta$. Clearly, a poor choice of $\theta$ might lead to situations with very large higher order cutoff 
effects\footnote{Note that the value $\theta=\pi/5$ chosen in \cite{Sint:1995ch} leads to very reduced cutoff effects for their
choice of BF. This would not be the case if this value was used with our BF.}.

It is remarkable that the value $\theta=\pi/2$, established in section \ref{subsec:theta} to obtain a condition number as small as possible,
also leads to a situation where higher order cutoff effects are highly suppressed.

A very similar picture is observed when considering any of the 2-index representations. 
Cutoff effects for the adjoint, symmetric and anti-symmetric representations are shown respectively 
in panels (b), (c) and (d) of Fig.~\ref{fig:fermionic_cutoff_effects}.
The smallness of the residual lattice artifacts is at first glance surprising, since they have previously 
been reported to be very large if particular care is not taken in the choice of BF \cite{Sint:2011gv,Karavirta:2012qd,Sint:2012ae}.
The magnitude of $\delta_{1,1}$ for the 2-index representations strongly depends on the angle $\theta$ in a very
similar way as it is shown in Fig.~\ref{fig:cutoff_effects_theta_su3_su6} for the fundamental representation. It is then possible to tune
$\theta$ to minimize cutoff effects without the need of modifying the BF \cite{Sint:2011gv,Karavirta:2012qd}. What is remarkable of the family of background 
fields proposed in this work is that for the fundamental, symmetric and antisymmetric representations, values of $\theta$ which lead to small
condition numbers also lead to small higher order lattice artifacts in the step scaling function. This is not true for the adjoint 
representation since, as discussed in section \ref{subsec:theta}, the condition number is minimized for $\theta=0$. 
It is also remarkable that cutoff effects for all the representations considered are minimized for the same value $\theta=\pi/2$.

\section{Matching the $\Lambda$ parameter to $\overline{\rm MS}$\label{sec:matching}}
In this section we calculate the relation $\Lambda_{\textrm{SF}}/\Lambda_{\overline{\textrm{MS}}}$
of $\Lambda$ parameters in our family of SF schemes and the $\overline{\textrm{MS}}$ scheme.
This relation is essential for obtaining the ratio $\Lambda_{\overline{\textrm{MS}}}/\sqrt{\sigma}$
from SF simulations. We provide numerical values of $\Lambda_{\textrm{SF}}/\Lambda_{\overline{\textrm{MS}}}$
for the pure gauge theories and for the theories with 2 fundamental fermions.
For completeness, we derive an expression (see Eq.~\ref{ratio_sf_ms}) for the ratio
$\Lambda_{\textrm{SF}}/\Lambda_{\overline{\textrm{MS}}}$ as a function of $N$, $N_{f}$
and the representation $\textrm{R}$, which might be useful also for future
BSM studies using the SF.

The $\Lambda$ parameter is a renormalization group invariant and scheme dependent quantity given by (in a generic scheme $X$) 
\begin{equation}
 \Lambda_{X}=\mu\left(b_{0}g_{X}^{2}\right)^{-\frac{b_{1}}{(2b_{0}^{2})}}e^{-\frac{1}{(2b_{0}g_{X}^{2})}}\exp\left\{-\int_{0}^{g_{X}}dx\left[\frac{1}{\beta_{X}(x)}+\frac{1}{b_{0}x^{3}}-\frac{b_{1}}{b_{0}^{2}x}\right]\right\}.
\label{lambda_X}
\end{equation}
It is a dimensionfull scale dynamically generated by the theory.

In subsection \ref{sec:results} we have performed the computation of the SF coupling\footnote{In the following we write explicitly a subindex with the coupling to indicate 
the scheme.} $\overline{g}_{\textrm{SF}}$ in
the Schr\"{o}dinger Functional scheme to one loop order in 
perturbation theory, i.e. we have calculated the
renormalized coupling as an expansion in terms of the bare coupling $g_{0}$
\begin{equation}
 \overline{g}_{\textrm{SF}}^{2}(L)=g_{0}^{2}+p_{1}(L/a)g_{0}^{4} + O(g_{0}^{6}),
\label{sf_bare}
\end{equation}
where, after doing a continuum extrapolation, 
\begin{equation}
  p_{1}(L/a)=r_{0}+2b_{0}\ln(L/a).
\end{equation}

To be able to compare the results at different values of $N$, we are interested in
a relation between $\alpha_{\textrm{SF}}=\overline{g}^{2}_{\textrm{SF}}/4\pi$ 
and some scheme where $N$
is only a parameter. For that we choose the usual 
$\overline{\textrm{MS}}$ scheme, defined at infinite volume and at high energies. 
The relation between the running coupling in the two schemes 
can be written as an expansion
\begin{equation}
 \alpha_{\overline{\textrm{MS}}}(s\mu)=\alpha_{\textrm{SF}}(\mu)+c_{1}(s)\alpha_{\textrm{SF}}(\mu)^{2}+c_{2}(s)\alpha_{\textrm{SF}}(\mu)^{3}+O(\alpha_{\textrm{SF}}^{4}),
\label{ms_sf}
\end{equation}
where $s$ is a scale parameter and $c_{i}(s)$ are the coefficients relating the
couplings in the two schemes at each order in perturbation theory.

The relation between the $\Lambda$ parameter in the 
SF and $\overline{\textrm{MS}}$ scheme is given by 
\begin{equation}
  \frac{\Lambda_{\textrm{SF}}}{\Lambda_{\overline{\textrm{MS}}}}=\exp\left\{-\frac{c_{1}(1)}{(8\pi b_{0})}\right\},
\label{lambdaSF_lambdaMS}
\end{equation}
where $c_{1}(1)$ 
 is the coefficient of the 1-loop relation~\eqref{ms_sf}. Note that
 Eq.~\eqref{lambdaSF_lambdaMS} is an exact relation even though
 it depends on the 1-loop coefficient relating the couplings in two
 different schemes. 

For determining the coefficient $c_{1}(s)$ in Eq.~\eqref{ms_sf}, we first use the known relation between $\alpha_{\overline{\textrm{MS}}}$ in the $\overline{\textrm{MS}}$
and the bare coupling $\alpha_{0}$ \cite{Luscher:1996ug,Sint:1995rb,Luscher:1996sc,DelDebbio:2008wb}, which at 1-loop is given by 
%\begin{equation}
% \alpha_{\overline{\textrm{MS}}}(s/a)=\alpha_{0}+d_{1}(s)\alpha_{0}^{2}+d_{2}(s)\alpha_{0}^{3}+O(\alpha_{0}^{4}).
%\label{ms_bare}
%\end{equation}
\begin{equation}
 \alpha_{\overline{\textrm{MS}}}(s/a)=\alpha_{0}+d_{1}(s)\alpha_{0}^{2}+O(\alpha_{0}^{3}).
\label{ms_bare}
\end{equation}
The 1-loop coefficient $d_{1}(s)$ is given for generic $N$ and fermionic representation $R$ by
\begin{equation}
 d_{1}(s) = d_{1,0} + N_{f}d_{1,1} -8\pi b_{0}\ln(s) 
\label{d1}
\end{equation}
where 
\begin{equation}
 d_{1,0}= -\frac{\pi}{2N}+k_{1}N \qquad \textrm{and}\qquad d_{1,1}=\widetilde{K}_{1}T_{R}.
\end{equation}
The coefficient $k_{1}$ of the gauge part is taken from \cite{Luscher:1996ug,Sint:1995rb,Luscher:1996sc}
and reads
\begin{equation}
  k_{1} = 2.135730074078457(2)
\label{k1}
\end{equation}
The coefficient $\widetilde{K}_{1}$ is a representation independent function 
of the tree level coefficient 
$c_{\textrm{SW}}^{(0)}$ given by
\begin{equation}
\widetilde{K}_{1}(c_{\textrm{SW}}^{(0)})=-0.1682888(2)+0.126838(2)c_{\textrm{SW}}^{(0)}-0.750048(2)(c_{\textrm{SW}}^{(0)})^{2} 
\label{K1}
\end{equation}
It was calculated in \cite{Weisz:1980pu,Christou:1998ws,Capitani:1994qn} for the fundamental 
representation and extended to
arbitrary representations in \cite{DelDebbio:2008wb}.

Combining Eqs.~\eqref{sf_bare} and \eqref{gbarexpansion}
we obtain the coefficient for the relation \eqref{ms_sf}, i.e.
\begin{equation}
 c_{1}(s) = d_{1,0}+N_{f}d_{1,1}  -4\pi(r_{0,0}+N_{f}r_{0,1}) -8\pi b_{0}\ln(s),
\end{equation}
with $r_{0,i}$ being the continuum coefficients in the series \eqref{eq:p1i_series}.

Knowing this, the relation between $\Lambda$ parameters in Eq.~\eqref{lambdaSF_lambdaMS}
can be given as a function of the parameters $N$, $N_{f}$ and $T_{R}$ 
and of the coefficients $r_{0,i}$ 
\begin{equation}
\frac{\Lambda_{\textrm{SF}}}{\Lambda_{\overline{\textrm{MS}}}}=\exp\left\{ \frac{3\pi^{2}/N - 6\pi( k_{1}N+\widetilde{K}_{1}T_{R}N_{f}) + 24\pi^{2}(r_{0,0}+N_{f}r_{0,1})}{11N-4T_{R}N_{f}} \right\}.
\label{ratio_sf_ms}
\end{equation}
Finally, in Table \ref{tab:lambdarel} we collect the values of the ratio of $\Lambda$ parameters for the schemes studied in this work and for the
pure gauge theory ($N_{f}=0$) and for 2 flavors of fundamental fermions
\footnote{Note that different choices of boundary phases or $\theta$ parameter
correspond to different choices of renormalization scheme and will hence
lead to different values for the ratio $\Lambda_{\textrm{SF}}/\Lambda_{\overline{\textrm{MS}}}$.}.
Ratios of lambda parameters
for 2 index representations can be recovered using Eq.~(\ref{ratio_sf_ms}) and the 
corresponding coefficients from Table~(\ref{tab:icoeff_fermion}). 

\begin{table}
\begin{center}
\begin{tabular}{rrr}
\hline
 $N$ & $\quad\left.\Lambda_{\textrm{SF}}/\Lambda_{\overline{\textrm{MS}}}\right|_{N_{f}=0}$& $\quad\left.\Lambda_{\textrm{SF}}/\Lambda_{\overline{\textrm{MS}}}\right|_{N_{f}=2}$\\
\hline
 $2$  &  $  0.44566597(1) $  &  $0.779492(3)$ \\
 $3$  &  $  0.48811256(1) $  &  $0.699183(2)$ \\
 $4$  &  $  0.503112529(5) $  & $0.654811(1)$ \\
 $5$  &  $  0.521195149(4) $  & $0.6426328(9)$\\
 $6$  &  $  0.539386422(6) $  & $0.6422183(8)$\\
 $7$  &  $  0.556975178(5) $  & $0.6470850(6)$\\
 $8$  &  $  0.573795805(3) $  & $0.6545895(6)$\\
 $9$  &  $  0.589843423(7) $  & $0.6634798(5)$\\
 $10$  &  $  0.60516382(7) $  & $0.6731035(4)$\\
 $11$  &  $  0.61981639(6) $  & $0.6830971(4)$\\
 $12$  &  $  0.63385977(6) $  & $0.6932473(3)$\\
\hline
\end{tabular}
\caption{Ratios between $\Lambda$ parameters in the SF and
  $\overline{\textrm{MS}}$ schemes, for the pure gauge theory and for 2 flavors
of fundamental fermions. \label{tab:lambdarel}}
\end{center}
\end{table}

\section{Conclusions\label{sec:conclusions}}
We have studied the Schr\"odinger functional boundary conditions and
the perturbative $\mathcal{O}(a)$ improvement for SU($N$) gauge
theories with general $N$. The improvement coefficient $c_t ^{(1,0)}$
is obtained also for all values of $N$. Additionally we provide the
matching between the SF and $\overline{\textrm{MS}}$ schemes for a wide range
of theories including fermions in various representations. This
enables a precision study of the coupling and the determination of
$\Lambda_{\overline{\rm MS}}$ in the large $N$ limit.

 The fermionic twisting angle $\theta$ is also studied and we found out that 
the value $\theta=\pi/2$ is a good compromise between the simulation
speed and the minimization of the $\mathcal{O}(a^2)$ lattice artifacts
in the perturbative 1-loop lattice step scaling function. 

\acknowledgments
This work was supported by the Danish National Research Foundation
DNRF:90 grant and by a Lundbeck Foundation Fellowship grant. TK is also
funded by the Danish Institute for Advanced Study. PV gratefully acknowledges
support by the EU under grant agreement number PITN-GA-2009-238353 (ITN STRONGnet)
and by the INFN progetto premiale ``SUMA''.
We thank Biagio Lucini and Stefan Sint for useful discussions. 

\appendix

\section{Details of the perturbative calculations}
\label{app:pertdetails}
In this appendix we provide details on the calculation for an arbitrary group
SU($N$) of the ghost and gauge contributions $h_{0}(L/a)$ and $h_{1}(L/a)$ 
to the 1 loop coupling (see Eqs. \ref{eq:hs} and \ref{eq:p10}). 
All the calculations are presented in a general
framework. The specific values of different variables with our choice
of background field and basis for SU($N$) generators are shown in
Appendix~\ref{app:basis}. In the following we will work in lattice
units i.e. the lattice spacing $a=1$. Additionally repeated Latin
indices $a,b,c,\ldots$ are {\emph{not}} summed over and repeated Greek
indices $\alpha,\beta,\gamma,\ldots$ are always summed over unless otherwise stated in the formula. Latin indices run from $1,2,3$ and Greek ones from $0,1,2,3$. 

The operators we are interested in are defined as
\be
\Delta_0\omega(x)&=&-D^*_{\mu}D_{\mu}\omega(x),\label{delta0}\\
\Delta_1 q_{\mu}(x)&=&-\lambda_0D_{\mu}D^*_{\nu}q_{\nu}(x)+\sum_{\nu\neq\mu}\left\{\cosh(a^2 G_{\mu\nu})\star [-D^*_{\nu}D_{\nu}q_{\mu}(x)+D^*_{\nu}D_{\mu}q_{\nu}(x)]\right.\nonumber\\
&&\left.-a^{-2}\sinh(a^2 G_{\mu\nu})\star [2 q_{\nu}(x)+a(D^*_{\nu}+D_{\nu})q_{\mu}(x)+a^2D^*_{\nu}D_{\mu}q_{\nu}(x)]\right\},\label{delta1}\\
\Delta_2 \psi(x)&=&\left[\left(D_{WD}+m_0\right)\gamma_5\right]^2\psi(x).\label{delta2}
\ee
There is no summation over $\mu$ in the r.h.s. of the Eq.~\eqref{delta1}. The star product in Eq.~\eqref{delta1} which maps an $N\times N$ matrix $M$ and an SU($N$) matrix $X$ to an SU($N$) matrix is defined as
\be
M\star X = \left(MX+XM^{\dag}\right)/2-\Tr\left(MX+XM^{\dag}\right)/(2 N).
\label{starproduct}
\ee
In Eq.~\eqref{delta2} the operator $D_{WD}$ is the same as in Eq.~\eqref{wilsondirac} with $c_{sw}=1$.

The first step is to find a suitable basis for the SU($N$)
generators. This is a basis that is invariant under the star product
defined in Eq.~\eqref{starproduct}. In practice we want to find
generators $X^a$ that satisfy 
\be\begin{array}{lcl}
\cosh G_{0k}\star X^a&=&\chi_a ^c X^a,\\
\sinh G_{0k}\star X^a&=&\chi_a ^s X^a,
\label{starcoefs}
\end{array}\ee
with arbitrary coefficients $\chi_a ^c$ and $\chi_a ^s$. The hyperbolic sine and cosine of the non-zero elements of the field strength tensor are
\be
\cosh G_{0k}=\cos\left[(C'_k-C_k)/L\right],\qquad \sinh G_{0k}=i\sin\left[(C'_k-C_k)/L\right].
\ee
A basis that satisfies Eq.~\eqref{starcoefs} for the non-diagonal generators are the ladder operators defined as
\be
\left(X^{a(j,k)}\right)_{nm}&=&-i/2\delta_{jn}\delta_{km},
\label{eq:nondiaggen}
\ee
where $n$ and $m$ are the matrix indices and $a(j,k)$ is the color index. The properties of $a(j,k)$ are given in Table~\ref{a table}. The generators $X^a$ are normalized as $\Tr\left[X^a X^b\right]=-\frac{1}{2}\delta^{a,b}$. The diagonal generators can be chosen in any way that satisfies Eq.~\eqref{starcoefs}. 

\begin{table}
\begin{tabular}{cccc}  
\hline
$a(j,k)$ & $\textrm{Range in } a$ & $\textrm{Range in } j$  &$\textrm{Range in } k$ \\
\hline\hline
 ~~ $(N - j/2) (j - 1) + (k - j)$ ~~ & $1,\ldots,N(N-1)/2$ &  ~~ $ 1,\ldots, N-1$ ~~ & ~~ $j+1,\ldots,N$ ~ ~ \\
\hline
~~ $ (j-1)(j/2-1) + k+N(N-1)/2$ ~~ & ~~ $N(N-1)/2+1,\ldots,N^2-N$ ~~ & $2,\ldots, N$ & $1,\ldots,j-1$ \\
\hline
\end{tabular}
\caption{The values of the color index $a(j,k)$ as a function of the dummy indices $j$ and $k$. When $1\leq a\leq (N^2-N)/2$  the generators $X^a$ have a non-zero element in the upper and for $(N^2-N)/2<a\leq N^2-N$ in the lower triangle.}
\label{a table}
\end{table}

%\be
%a(j,k)=\left\{\begin{array}{lll}(N - j/2) (j - 1) + (k - j),& \begin{array}{l} j\in\{1,\ldots, N-1\},\\ k\in\{j+1,\ldots,N\},\end{array}& 1\leq a\leq (N^2-N)/2,\\(j-1) (j - 2) + k,& \begin{array}{l} j\in\{2,\ldots, N\},\\ k\in\{1,\ldots,j\},\end{array}& (N^2-N)/2<a\leq N^2-N,
%\end{array}\right.
%\ee

The boundary conditions generate a background field $V_\mu(x)$ defined
in Eqs.~\eqref{background field}, \eqref{Bdefinition1} and \eqref{Bdefinition2} which enters the covariant derivatives
\be\begin{array}{lcl}
D_{\mu}q(x)&=&\left[V_\mu(x)q(x+\hat{\mu})V^{-1}_\mu(x)-q(x) \right],\\
D^*_{\mu}q(x)&=&\left[q(x)-V^{-1}_\mu(x-\hat{\mu})q(x-\hat{\mu})V_\mu(x-\hat{\mu})\right],
\end{array}\label{eq:covder}
\ee
and through them to the operators $\Delta_s$. The next step is to calculate the covariant derivatives with the background field $V_\mu(x)$ when $q(x)=q_a(x)X^a$ is proportional to a generator $X^a$. The covariant derivatives can then be written in a general form
\be
D_{\mu}q(x)&=&\left\{\begin{array}{lcl}\left[q_a(x+\hat{\mu})-q_a(x)\right]X^a, &\textrm{if} &\mu=0,\\ \left[\exp(i f_a)q_a(x+\hat\mu)+q_a(x)\right]X^a,&\textrm{if} &\mu>0,\end{array}\right. \\
D^*_{\mu}q(x)&=&\left\{\begin{array}{lcl}\left[q_a(x)-q_a(x-\hat{\mu})\right]X^a, &\textrm{if} &\mu=0,\\ \left[q_a(x)-\exp(-i f_a)q_a(x-\hat\mu)\right]X^a,&\textrm{if} &\mu>0,\end{array}\right.
\ee
where
\be
[B_k(x),X^a]=i f_a(t) X^a,
\ee
and $t$ is the time component of the four vector $x=(t,\vec{x})$.
The operators $\Delta_0$ and $\Delta_1$ can now be decomposed to color subspaces according to the basis selected. The operators are also invariant under spatial translations and thus the determinants can be written as
\be
\det\Delta_s=\prod_a\prod_{p}\det\Delta_s|_{(p,a)},\qquad s=0,1,
\ee
where 
\be
p=2\pi n/L,\qquad n_k\in\mathbb{Z},\qquad -L/2<n_k\leq L/2,
\ee
is the three momentum. % and $a$ is the colour index.

Next we will show how one can calculate the determinant of operators $\Delta_s$. In \cite{Luscher:1992an} it has been shown that for an operator $\Delta$ that satisfies
\be
\Delta\psi(t)=A(t)\psi(t+1)+B(t)\psi(t)+C(t)\psi(t-1),
\label{deltarecursion}
\ee 
for matrices $A$, $B$ and $C$ and an eigenvalue equation
\be
\left\{\begin{array}{ll}(\Delta-\xi)\psi(t)=0,&\qquad t>0,\\
\psi(0)=\psi(L)=0, & \end{array}\right.
\label{eigeneq}
\ee
there exists a matrix $M(\xi)$ such that
\be
\psi(L)=M(\xi)\psi(1)=0.
\ee
The determinant of $\Delta$ is then given by
\be
\det\Delta=\det M(0)\prod_{t=1}^{L-1} \det[-A(t)].
\label{detequation}
\ee
We will next use these properties of the $\Delta_s$ operators.

Since the operator $\Delta_0$ is invariant under spatial translations and constant diagonal gauge transformations its eigenfunctions are of the form
\be
\omega_a(x)=\psi_a(t)e^{ipx}X^a.
\ee
Operating with $\Delta_0$ on $\omega_a(x)$ we get
\be
\Delta_0\omega_a(x)=\left[A \psi_a(t+1)+B_a(t)\psi_a(t)+C \psi_a(t-1)\right]e^{ipx}X^a,\label{recursion1}
\ee
with $A=C=-1$ and
\be
B_a(t)=8-2\sum_{k=1}^3\cos\left[p_k+f_a(t)\right].
\ee
Clearly the operator $\Delta_0$ is similar to the operator in Eq.~\eqref{deltarecursion} and thus the strategy shown can be used. Using the Eq.~\eqref{eigeneq} with $\xi=0$ i.e.
\be
\Delta_0\psi_a(t)=0,\qquad 0\leq t <L
\ee
we get a recursion relation for $\psi_a(t)$ with initial values $\psi_a(0)=0$, $\psi_a(1)=1$ which is 
\be
\psi_a(2)&=&B_a(1),\\
\psi_a(t+1)&=&B_a(t)\psi_a(t)-\psi_a(t-1),\qquad t\geq 2.
\ee
According to Eq.~\eqref{detequation} the determinant is then
\be
\det\Delta_0|_{(p,a)}=\psi_a(L).
\ee  

We will then move on to the more challenging case of $\Delta_1$. The eigenfunctions of the operator $\Delta_1$ have the general form
\be
q^a _{\mu}(x)=R^a _{\mu\nu}(t)\psi^a _{\nu}(t) e^{ipx}X^a,  
\label{efdelta1}
\ee
where normalization\footnote{Adding $R^a _{\mu\nu}$ ensures that the matrices  $A^a _{\mu\nu}(t)$, $B^a _{\mu\nu}(t)$ and $C^a _{\mu\nu}(t)$ in the recursion relation are real.} matrix $R^a _{\mu\nu}(t)$ is a diagonal $4\times 4$ matrix with
\be
R^a _{00}(t)=-i,\qquad \qquad R^a _{kk}(t)=e^{i(p_k+f_a(t))/2},\quad k=1,2,3. 
\ee
Again we can operate with $\Delta_1$ on the eigenfunction Eq.~\eqref{efdelta1} which yields
\be
\Delta_1 q^a _{\mu}(x)=R^a_{\mu\nu}(t)\left[A^a _{\nu\rho}(t)\psi^a _{\rho}(t+1)+B^a _{\nu\rho}(t)\psi^a _{\rho}(t)+C^a _{\nu\rho}(t)\psi^a _{\rho}(t-1)\right] e^{ipx}X^a,\label{recursion2}
\ee
where the matrices $A^a _{\mu\nu}(t)$, $B^a _{\mu\nu}(t)$ and $C^a _{\mu\nu}(t)$ are 
\be\begin{array}{l}
A^a _{\mu\nu}(t)=\left\{\begin{array}{lcl} A^a _{00}(t)&=&-\lambda_0, \\ A^a _{kl}(t) &=&-N^a \delta_{k,l},\\ A^a _{0k}(t) &=& \lambda_0 s^a _k(t+1)-N^a s^a _k(t), \\ A^a _{k0}(t) &=& 0,\end{array}\right.\\ \\
B^a _{\mu\nu}(t)=\left\{\begin{array}{lcl} B^a _{00}(t)&=&2\lambda_0+\sum_{k=1} ^3 s^a _k(t)\left(\chi_a ^c s^a _k(t)-i \chi_a ^s c^a _k(t)\right), \\ B^a _{kl}(t) &=&(\lambda_0-1)s^a _k(t)s^a _l(t)+\delta_{k,l}(2 \chi_a ^c+\sum_{n=1} ^3(s^a _n(t))^2 )
,\\ B^a _{0k}(t) &=& \chi_a ^c s^a _k(t)-i \chi_a ^s c^a _k(t)-\lambda_0 s^a _k(t), \\ B^a _{k0}(t) &=& B^a _{0k}(t),\end{array}\right.\\ \\
C^a _{\mu\nu}(t)=A^a _{\nu\mu}(t-1).\end{array}
\ee
We have used the following short handed notation
\be
c^a _k(t)&=&2\cos\left[(p_k+f_a(t))/2\right],\\
s^a _k(t)&=&2\cos\left[(p_k+f_a(t))/2\right],\\
N^a&=&(\chi_a ^c-\chi_a ^s)\exp\left[i(f_a(t+1)-f_a(t))/2\right].
\ee
The operator $\Delta_1$ is also similar to the case in Eq.~\eqref{deltarecursion} and the same strategy can again be exploited. Additionally the boundary conditions of $\psi^a _{\mu}(t)$ in Eq.~\eqref{recursion2} are
\be
\psi^a _{0}(-1)&=&\partial^*\psi^a _{0}(L)=\psi^a _k(0)=\psi^a _k(L)=0, \qquad \vec{p}=0 \bigwedge a>N^2-N,\label{boundary p0}\\
\partial^*\psi^a _{0}(0)&=&\partial^*\psi^a _{0}(L)=\psi^a _k(0)=\psi^a _k(L)=0, \qquad \textrm{else}.\label{boundary else} 
\ee

With this we can first calculate the determinant of $\Delta_1$ in the more general case where the boundary conditions are given by Eq.~\eqref{boundary else}. Setting $\xi=0$ in Eq.~\eqref{eigeneq} we get
\be
\Delta_1\psi^a _0(t)&=&0,\qquad 0\leq t< L,\label{recrel1}\\
\Delta_1\psi^a _k(t)&=&0,\qquad 0< t< L
\label{recrel2}
\ee 
With the help of these equations we find $F^a _{\mu\nu}(t)$ which has the property 
\be
\psi^a _{\mu}(t)=F^a _{\mu\nu}(t)v^a _{\nu},
\ee
where 
\be
v^a _{\nu}=\left(\begin{array}{c} \psi^a _0(0)\\ \psi^a _k(1)\end{array}\right),
\label{boundary v}
\ee
are the first nonzero components of $\psi^a _{\mu}(t)$. The matrix $F^a _{\mu\nu}(t)$ is
\be\begin{array}{l}
F^a _{\mu\nu}(1)=\left\{\begin{array}{ll} -\left[B^a _{00}(0)+C^a _{00}(0)\right]/A^a _{00}(0), &\qquad \mu=\nu=0, \\ -A^a _{0k}(0)/A^a _{00}(0), &\qquad \mu=0\bigwedge\nu=k\neq 0, \\ 0, &\qquad \mu\neq 0\bigwedge \nu=0, \\ \delta_{k,l}, &\qquad \mu=k\neq 0\bigwedge \nu=l\neq 0, \end{array}\right.\\ \\
F^a _{\mu\nu}(2)=-\left(A^a _{\mu\rho}(1)\right)^{-1}\left[B^a _{\rho\sigma}(1)F^a _{\sigma\nu}(1)+C^a _{\rho\sigma}(1)P_{\sigma\nu} \right],\\ \\
F^a _{\mu\nu}(t+1)=-\left(A^a _{\mu\rho}(t)\right)^{-1}\left[B^a _{\rho\sigma}(t)F^a _{\sigma\nu}(t)+C^a _{\rho\sigma}(t)F^a_{\sigma\nu}(t-1) \right],\qquad t \geq 2,
\end{array}\ee
where the projection operator $P_{\mu\nu}$ is
\be
P_{\mu\nu}=\left\{\begin{array}{ll} 1, &\qquad \mu=\nu=0, \\ 0, &\qquad \textrm{else}.\end{array}\right.  
\ee
With $F^a _{\mu\nu}(t)$ we will be able to construct a matrix $M^a _{\mu\nu}$ that couples $v^a _{\mu}$ from Eq.~\eqref{boundary v} and the boundary condition Eq.~\eqref{boundary else} at $t=L$
\be
\left(\begin{array}{cc}\partial^*\psi^a _0(L)\\ \psi^a _k(L)\end{array}\right)=M^a _{\mu\nu}v^a _{\nu}.
\label{defM}
\ee
This matrix $M^a _{\mu\nu}$ turns out to be 
\be
M^a _{\mu\nu}=F^a _{\mu\nu}(L)-P_{\mu\rho}F^a _{\rho\nu}(L-1),
\ee
and the determinant of $\Delta_1$ in this subspace according to Eq.~\eqref{detequation} is
\be
\det\Delta_1|_{(p,a)}=\det\left[M^a _{\mu\nu} \lambda_0 ^L (N^a)^{3(L-1)}\right].
\ee

We can then move on to the case of $\Delta_1$ where $a>N^2-N$ i.e. for diagonal generators $X^a$ and when $\vec{p}=0$. In this case the boundary conditions are given by Eq.~\eqref{boundary p0} and $\psi^a _0(t)$ and $\psi^a _k(t)$ 
components decouple since the matrices $A^a _{\mu\nu}(t)$, $B^a _{\mu\nu}(t)$ and $C^a _{\mu\nu}(t)$ are diagonal
\be
A^a _{00}(t)&=&-\lambda_0,\\
A^a _{kk}(t)&=&-\chi_a ^c,\quad k=1,2,3,\\
A^a _{\mu\nu}(t)&=&C^a _{\mu\nu}(t)=1/2 B^a _{\mu\nu}(t).
\ee
Again using the Eq.~\eqref{recrel1} and Eq.~\eqref{recrel2} with the boundary conditions \eqref{boundary p0} we find that 
\be
\psi^a _0(L)=(L+1)\psi^a _0(0),\qquad \psi^a _k(L)=L\psi^a _k(1),
\ee
and then we can write down the matrix $M$ that couples the first nonzero components of $\psi^a _{\mu}(t)$ and the boundary condition \eqref{boundary p0} at $t=L$ as in Eq.~\eqref{defM}. Now the matrix $M$ is diagonal with entries
\be
M_{00}=1, \qquad  \qquad M_{kk}=L,\quad k=1,2,3.
\ee
According to the Eq.~\eqref{detequation} the contribution to the determinant of $\Delta_1$ is then
\be
\det\Delta_1|_{(p=0,a>N^2-N)}=\lambda_0 ^L L^3 (\chi_a ^c )^{3(L-1)}.
\ee

Now we are ready to present the value of the pure gauge part of the 1-loop coefficient $p_{1,0}(L/a)$ in the SF coupling which is
\be
p_{1,0}(L/a)&=&h_0(L/a)-1/2h_1(L/a)\nonumber\\
&=&\frac{1}{\kappa}\sum_{p,a,s}\frac{\partial}{\partial \eta}\left[\ln\det\Delta_0|_{(p,a)}-1/2\ln\det\Delta_1|_{(p,a)}\right]\nonumber\\
&=&\frac{1}{\kappa}\left\{\sum_{a=1}^{N^2-N}\sum_{p}\left[\frac{\psi_a'(L)}{\psi_a(L)}-\frac{3(L-1)}{2}\frac{(N^a)'}{N^a}-\frac{1}{2}M_{\mu\nu} ^{(-1)}M_{\nu\mu}' \right]\right.\nonumber\\
&&\quad\left.-\frac{1}{2}\sum_{a=N^2-N+1} ^{N^2-1}\left[3(L-1)\frac{(\chi_a ^c)'}{\chi_a ^c}+\sum_{p\neq 0}\left(3(L-1)\frac{(N^a)'}{N^a}+M_{\mu\nu} ^{(-1)}M_{\nu\mu}' \right)    \right] \right\},
\ee 
where prime indicates partial derivative w.r.t. the parameter $\eta$, $M_{\mu\nu}^{(-1)}$ is the inverse matrix of $M_{\mu\nu}$ and the normalization  $\kappa$ is defined in Eq.~\eqref{kappa}.

\section{Chosen basis for the diagonal generators and the values of the coefficients which depend on the background field}
\label{app:basis}
In Appendix~\ref{app:pertdetails} we showed how the 1-loop coupling can be calculated for a generic background field and basis of generators. In here we will specify the 
basis that we have selected as 
well as the values of the coefficients $\chi_a ^c$, $\chi_a ^s$ and $f_a(t)$.

We have chosen a basis given by
\be
\begin{array}{lcl}
X^{N^2-N+b}_{kk}&=&\frac{i}{\sqrt{2b(b+1)}}\left[-b\delta_{b+2,k}+\sum_{j=1}^{b}\delta_{j+1,k}    \right],\quad(N>3),\\
X^{N^2-2}_{kk}&=&i/2(\delta_{k,1}-\delta_{k,N}),\\
X^{N^2-1}_{kk}&=&\frac{i}{\sqrt{N(N-2)}}\left[-(N-2)(\delta_{k,1}+\delta_{k,N})+\sum_{j=2}^{N-1}\delta_{j,k}\right],
\end{array}
\ee
for the diagonal generators of SU($N$), and by Eq.(\ref{eq:nondiaggen}) 
for the non diagonal ones. 
With this choice the coefficients $\chi_a ^c$ and $\chi_a ^s$ from Eq.~\eqref{starcoefs} are
\be
\chi_{a(i,j)} ^c&=&\left\{\begin{array}{cc}1/2\left[\cos \zeta(i)+\cos \zeta(j)\right],&\quad 1\leq a(i,j)\leq N^2-N,\\ 0, & a(i,j)>N^2-N,\end{array}\right.\\
\chi_{a(i,j)} ^a&=&\left\{\begin{array}{cc}1/2\left[\sin \zeta(i)-\sin \zeta(j)\right],&\quad 1\leq a(i,j)\leq N^2-N\\ 0, & a(i,j)>N^2-N.\end{array}\right.
\ee
where
\be
\zeta(i)= \frac{2 \eta}{
      L^2} \left(\delta_{1, i} + \delta_{N, i} - 
       \frac{2}{N - 2} \sum_{k=2}^{N - 1} \delta_{k, i}\right)+ \frac{\pi}{
       N L^2} \left((N - 2)( \delta_{1, i} + \delta_{N,i}) - 2 \sum_{k=2}^{N - 1} \delta_{k, i}\right)
\ee

The coefficients $f_a(t)$ are\footnote{This expression holds for $N\geq 3$. In the case when $3\leq N<5$ it must be noted that the sum over $b$ does not exist and thus the first term does not contribute.}
\be
f_{a(i,j)}(t)&=&\frac{\pi}{2 L N}  \sum _{b=2}^{N-3} \left[\sum _{r=2}^{b+1} \left(\delta_{i,r}-\delta_{j,r}\right)-b
   \left(\delta_{i,b+2}-\delta_{j,b+2}\right)\right]\nonumber\\
   &&+\frac{(L-2 t) (2 \eta  N+\pi  (N-2))}{L^2 (N-2) N}
   \left[\sum _{r=2}^{N-1} \left(\delta_{r,i}-\delta_{r,j}\right)+\frac{ (N-2) }{2}\left(\delta_{N,j}-\delta_{1,i}\right)\right]\nonumber\\
   &&+\frac{\pi}{2 L N}  \left[\delta_{i,2}-\delta_{i,3}-\delta_{j,2}+\delta_{j,3}+ (2 N-1)\left(\delta_{1,i}+\delta_{N,j}\right)\right],
\ee
for $a\leq N^2-N$ and $f_a=0$ for $a>N^2-N$.

Additionally the normalization in Eq.~\eqref{kappa} depends on the chosen background field. For our choice it is 
\be
\kappa=24L^2\left[\sin\left(((1-2/N)\pi+2\eta)/L^2\right)+\sin\left(2/L^2(\pi/N+2\eta/(N-2))\right)\right].
\ee


\begin{thebibliography}{30}
  %\cite{Luscher:1992an}
\bibitem{Luscher:1992an}
  M.~Luscher, R.~Narayanan, P.~Weisz and U.~Wolff,
  %``The Schrodinger functional: A Renormalizable probe for nonAbelian gauge
  %theories,''
  Nucl.\ Phys.\  B {\bf 384}, 168 (1992)
  [arXiv:hep-lat/9207009].
  %%CITATION = NUPHA,B384,168;%%

%\cite{Luscher:1992zx}
\bibitem{Luscher:1992zx} 
  M.~Luscher, R.~Sommer, U.~Wolff and P.~Weisz,
  %``Computation of the running coupling in the SU(2) Yang-Mills theory,''
  Nucl.\ Phys.\ B {\bf 389}, 247 (1993)
  [hep-lat/9207010].
  %%CITATION = HEP-LAT/9207010;%%
  %128 citations counted in INSPIRE as of 25 Aug 2014


  %\cite{Luscher:1993gh}
\bibitem{Luscher:1993gh}
  M.~Luscher, R.~Sommer, P.~Weisz and U.~Wolff,
  %``A Precise determination of the running coupling in the SU(3) Yang-Mills
  %theory,''
  Nucl.\ Phys.\  B {\bf 413}, 481 (1994)
  [arXiv:hep-lat/9309005].
  %%CITATION = NUPHA,B413,481;%%

%\cite{DellaMorte:2004bc}
\bibitem{DellaMorte:2004bc} 
  M.~Della Morte {\it et al.}  [ALPHA Collaboration],
  %``Computation of the strong coupling in QCD with two dynamical flavors,''
  Nucl.\ Phys.\ B {\bf 713}, 378 (2005)
  [hep-lat/0411025].
  %%CITATION = HEP-LAT/0411025;%%
  %129 citations counted in INSPIRE as of 25 Aug 2014

%\cite{Lucini:2008vi}
\bibitem{Lucini:2008vi} 
  B.~Lucini and G.~Moraitis,
  %``The Running of the coupling in SU(N) pure gauge theories,''
  Phys.\ Lett.\ B {\bf 668}, 226 (2008)
  [arXiv:0805.2913 [hep-lat]].
  %%CITATION = ARXIV:0805.2913;%%
  %14 citations counted in INSPIRE as of 03 Jun 2014

%\cite{Appelquist:2007hu}
\bibitem{Appelquist:2007hu} 
  T.~Appelquist, G.~T.~Fleming and E.~T.~Neil,
  %``Lattice study of the conformal window in QCD-like theories,''
  Phys.\ Rev.\ Lett.\  {\bf 100}, 171607 (2008)
  [Erratum-ibid.\  {\bf 102}, 149902 (2009)]
  [arXiv:0712.0609 [hep-ph]].
  %%CITATION = ARXIV:0712.0609;%%
  %219 citations counted in INSPIRE as of 25 Aug 2014

%\cite{Hietanen:2009az}
\bibitem{Hietanen:2009az} 
  A.~J.~Hietanen, K.~Rummukainen and K.~Tuominen,
  %``Evolution of the coupling constant in SU(2) lattice gauge theory with two adjoint fermions,''
  Phys.\ Rev.\ D {\bf 80}, 094504 (2009)
  [arXiv:0904.0864 [hep-lat]].
  %%CITATION = ARXIV:0904.0864;%%
  %143 citations counted in INSPIRE as of 25 Aug 2014

%\cite{Karavirta:2011zg}
\bibitem{Karavirta:2011zg} 
  T.~Karavirta, J.~Rantaharju, K.~Rummukainen and K.~Tuominen,
  %``Determining the conformal window: SU(2) gauge theory with N_f = 4, 6 and 10 fermion flavours,''
  JHEP {\bf 1205}, 003 (2012)
  [arXiv:1111.4104 [hep-lat]].
  %%CITATION = ARXIV:1111.4104;%%
  %26 citations counted in INSPIRE as of 25 Aug 2014

%\cite{Bursa:2009we}
\bibitem{Bursa:2009we} 
  F.~Bursa, L.~Del Debbio, L.~Keegan, C.~Pica and T.~Pickup,
  %``Mass anomalous dimension in SU(2) with two adjoint fermions,''
  Phys.\ Rev.\ D {\bf 81}, 014505 (2010)
  [arXiv:0910.4535 [hep-ph]].
  %%CITATION = ARXIV:0910.4535;%%

%\cite{Bursa:2010xn}
\bibitem{Bursa:2010xn} 
  F.~Bursa, L.~Del Debbio, L.~Keegan, C.~Pica and T.~Pickup,
  %``Mass anomalous dimension in SU(2) with six fundamental fermions,''
  Phys.\ Lett.\ B {\bf 696}, 374 (2011)
  [arXiv:1007.3067 [hep-ph]].
  %%CITATION = ARXIV:1007.3067;%%

%\cite{DeGrand:2010na}
\bibitem{DeGrand:2010na} 
  T.~DeGrand, Y.~Shamir and B.~Svetitsky,
  %``Running coupling and mass anomalous dimension of SU(3) gauge theory with two flavors of symmetric-representation fermions,''
  Phys.\ Rev.\ D {\bf 82}, 054503 (2010)
  [arXiv:1006.0707 [hep-lat]].
  %%CITATION = ARXIV:1006.0707;%%
  %65 citations counted in INSPIRE as of 25 Aug 2014

%\cite{DeGrand:2012qa}
\bibitem{DeGrand:2012qa} 
  T.~DeGrand, Y.~Shamir and B.~Svetitsky,
  %``SU(4) lattice gauge theory with decuplet fermions: Schrodinger functional analysis,''
  Phys.\ Rev.\ D {\bf 85}, 074506 (2012)
  [arXiv:1202.2675 [hep-lat]].
  %%CITATION = ARXIV:1202.2675;%%
  %16 citations counted in INSPIRE as of 10 Jun 2014

%\cite{Hayakawa:2010yn}
\bibitem{Hayakawa:2010yn} 
  M.~Hayakawa, K.-I.~Ishikawa, Y.~Osaki, S.~Takeda, S.~Uno and N.~Yamada,
  %``Running coupling constant of ten-flavor QCD with the Schr\'odinger functional method,''
  Phys.\ Rev.\ D {\bf 83}, 074509 (2011)
  [arXiv:1011.2577 [hep-lat]].
  %%CITATION = ARXIV:1011.2577;%%
  %37 citations counted in INSPIRE as of 25 Aug 2014

%\cite{Heller:1997vh}
\bibitem{Heller:1997vh} 
  U.~M.~Heller,
  %``The Schrodinger functional running coupling with staggered fermions and its application to many flavor QCD,''
  Nucl.\ Phys.\ Proc.\ Suppl.\  {\bf 63}, 248 (1998)
  [hep-lat/9709159].
  %%CITATION = HEP-LAT/9709159;%%
  %16 citations counted in INSPIRE as of 25 Aug 2014

  %\cite{Sint:1995ch}
\bibitem{Sint:1995ch}
  S.~Sint and R.~Sommer,
  %``The Running coupling from the QCD Schrodinger functional: A One loop
  %analysis,''
  Nucl.\ Phys.\  B {\bf 465}, 71 (1996)
  [arXiv:hep-lat/9508012].
  %%CITATION = NUPHA,B465,71;%%
  
%\cite{Luscher:1996vw}
\bibitem{Luscher:1996vw}
  M.~Luscher and P.~Weisz,
  %``O(a) improvement of the axial current in lattice QCD to one-loop order  of
  %perturbation theory,''
  Nucl.\ Phys.\  B {\bf 479}, 429 (1996)
  [arXiv:hep-lat/9606016].
  %%CITATION = NUPHA,B479,429;%%
  
    %\cite{Luscher:1996sc}
\bibitem{Luscher:1996sc}
  M.~Luscher, S.~Sint, R.~Sommer and P.~Weisz,
  %``Chiral symmetry and O(a) improvement in lattice QCD,''
  Nucl.\ Phys.\  B {\bf 478}, 365 (1996)
  [arXiv:hep-lat/9605038].
  %%CITATION = NUPHA,B478,365;%%
  
 %\cite{Sheikholeslami:1985ij}
\bibitem{Sheikholeslami:1985ij}
  B.~Sheikholeslami and R.~Wohlert,
  %``Improved Continuum Limit Lattice Action For QCD With Wilson Fermions,''
  Nucl.\ Phys.\  B {\bf 259}, 572 (1985).
  %%CITATION = NUPHA,B259,572;%%

%\cite{Lucini:2012gg}
\bibitem{Lucini:2012gg} 
  B.~Lucini and M.~Panero,
  %``SU(N) gauge theories at large N,''
  Phys.\ Rept.\  {\bf 526}, 93 (2013)
  [arXiv:1210.4997 [hep-th]].
  %%CITATION = ARXIV:1210.4997;%%
  %68 citations counted in INSPIRE as of 03 Jun 2014
  
 


%\cite{Narayanan:2006rf}
\bibitem{Narayanan:2006rf} 
  R.~Narayanan and H.~Neuberger,
  %``Infinite N phase transitions in continuum Wilson loop operators,''
  JHEP {\bf 0603}, 064 (2006)
  [hep-th/0601210].
  %%CITATION = HEP-TH/0601210;%%
  %43 citations counted in INSPIRE as of 27 Jun 2014

%\cite{Luscher:2010iy}
\bibitem{Luscher:2010iy} 
  M.~Luscher,
  %``Properties and uses of the Wilson flow in lattice QCD,''
  JHEP {\bf 1008}, 071 (2010)
  [arXiv:1006.4518 [hep-lat]].
  %%CITATION = ARXIV:1006.4518;%%
  %83 citations counted in INSPIRE as of 03 Jun 2014


%\cite{Luscher:2011bx}
\bibitem{Luscher:2011bx} 
  M.~Luscher and P.~Weisz,
  %``Perturbative analysis of the gradient flow in non-abelian gauge theories,''
  JHEP {\bf 1102}, 051 (2011)
  [arXiv:1101.0963 [hep-th]].
  %%CITATION = ARXIV:1101.0963;%%
  %33 citations counted in INSPIRE as of 03 Jun 2014


%\cite{Ramos:2013gda}
\bibitem{Ramos:2013gda} 
  A.~Ramos,
  %``The gradient flow in a twisted box,''
  arXiv:1308.4558 [hep-lat].
  %%CITATION = ARXIV:1308.4558;%%
  %6 citations counted in INSPIRE as of 03 Jun 2014

  
%\cite{Fritzsch:2013je}
\bibitem{Fritzsch:2013je} 
  P.~Fritzsch and A.~Ramos,
  %``The gradient flow coupling in the Schrödinger Functional,''
  JHEP {\bf 1310}, 008 (2013)
  [arXiv:1301.4388 [hep-lat]].
  %%CITATION = ARXIV:1301.4388;%%
  %23 citations counted in INSPIRE as of 27 May 2014

%\cite{Ramos:2014}
\bibitem{Ramos:2014} 
  A.~Ramos,
  Plenary talk in Lattice 2014.

%\cite{Rantaharju:2013bva}
\bibitem{Rantaharju:2013bva} 
  J.~Rantaharju,
  %``The Gradient Flow Coupling in Minimal Walking Technicolor,''
  arXiv:1311.3719 [hep-lat].
  %%CITATION = ARXIV:1311.3719;%%
  %1 citations counted in INSPIRE as of 03 Jun 2014
  
%\cite{Narayanan:2003fc}
\bibitem{Narayanan:2003fc} 
  R.~Narayanan and H.~Neuberger,
  %``Large N reduction in continuum,''
  Phys.\ Rev.\ Lett.\  {\bf 91}, 081601 (2003)
  [hep-lat/0303023].
  %%CITATION = HEP-LAT/0303023;%%
  %85 citations counted in INSPIRE as of 03 Jun 2014


%\cite{Sannino:2004qp}
\bibitem{Sannino:2004qp} 
  F.~Sannino and K.~Tuominen,
  %``Orientifold theory dynamics and symmetry breaking,''
  Phys.\ Rev.\ D {\bf 71}, 051901 (2005)
  [hep-ph/0405209].
  %%CITATION = HEP-PH/0405209;%%
  %289 citations counted in INSPIRE as of 25 Aug 2014
  
  %\cite{Karavirta:2013qqa}
\bibitem{Karavirta:2013qqa} 
  T.~Karavirta, A.~Hietanen and P.~Vilaseca,
  %``Schr\"odinger functional boundary conditions and improvement of
  %the SU($N$) pure gauge action for $N>3$,''
  PoS(LATTICE 2013)328 
  arXiv:1311.0405 [hep-lat].
  %%CITATION = ARXIV:1311.0405;%%
 
%\cite{Sint:1995rb}
\bibitem{Sint:1995rb} 
  S.~Sint,
  %``One loop renormalization of the QCD Schrodinger functional,''
  Nucl.\ Phys.\ B {\bf 451}, 416 (1995)
  [hep-lat/9504005].
  %%CITATION = HEP-LAT/9504005;%%

 %\cite{Wohlert:1987rf}
\bibitem{Wohlert:1987rf}
  R.~Wohlert, %{\em Improved Continuum Limit Lattice Action For Quarks},
  %%CITATION = 
  DESY87/069%%  
  
  %\cite{Karavirta:2011mv}
\bibitem{Karavirta:2011mv} 
  T.~Karavirta, A.~Mykkanen, J.~Rantaharju, K.~Rummukainen and K.~Tuominen,
  %``Nonperturbative improvement of SU(2) lattice gauge theory with adjoint or fundamental flavors,''
  JHEP {\bf 1106}, 061 (2011)
  [arXiv:1101.0154 [hep-lat]].
  %%CITATION = ARXIV:1101.0154;%%
  %22 citations counted in INSPIRE as of 11 Jun 2014
  
%\cite{Luscher:1996ug}
\bibitem{Luscher:1996ug} 
  M.~Luscher, S.~Sint, R.~Sommer, P.~Weisz and U.~Wolff,
  %``Nonperturbative O(a) improvement of lattice QCD,''
  Nucl.\ Phys.\ B {\bf 491}, 323 (1997)
  [hep-lat/9609035].
  %%CITATION = HEP-LAT/9609035;%%
  
  %\cite{Sint:2012ae}
\bibitem{Sint:2012ae} 
  S.~Sint and P.~Vilaseca,
  %``Lattice artefacts in the Schr\'odinger Functional coupling for strongly interacting theories,''
  PoS LATTICE {\bf 2012}, 031 (2012)
  [arXiv:1211.0411 [hep-lat]].
  %%CITATION = ARXIV:1211.0411;%%
  %1 citations counted in INSPIRE as of 18 Oct 2013
  
  %\cite{Karavirta:2012qd}
\bibitem{Karavirta:2012qd}
  T.~Karavirta, K.~Tuominen and K.~Rummukainen,
  %``Perturbative Improvement of the Schrodinger Functional for Lattice Strong Dynamics,''
  Phys.\ Rev.\ D {\bf 85} (2012) 054506
  [arXiv:1201.1883 [hep-lat]].
  %%CITATION = ARXIV:1201.1883;%%
  %5 citations counted in INSPIRE as of 24 Jul 2014

%\cite{Sint:2011gv}
\bibitem{Sint:2011gv}
  S.~Sint and P.~Vilaseca,
  %``Perturbative lattice artefacts in the SF coupling for technicolor-inspired models,''
  PoS LATTICE {\bf 2011} (2011) 091
  [arXiv:1111.2227 [hep-lat]].
  %%CITATION = ARXIV:1111.2227;%%
  %5 citations counted in INSPIRE as of 24 Jul 2014

%\cite{Bode:1998hd}
\bibitem{Bode:1998hd} 
  A.~Bode {\it et al.}  [Alpha Collaboration],
  %``Two loop computation of the Schrodinger functional in pure SU(3) lattice gauge theory,''
  Nucl.\ Phys.\ B {\bf 540}, 491 (1999)
  [hep-lat/9809175].
  %%CITATION = HEP-LAT/9809175;%%
  %25 citations counted in INSPIRE as of 25 Aug 2014

%\cite{Bode:1999sm}
\bibitem{Bode:1999sm} 
  A.~Bode {\it et al.}  [ALPHA Collaboration],
  %``Two loop computation of the Schrodinger functional in lattice QCD,''
  Nucl.\ Phys.\ B {\bf 576}, 517 (2000)
  [Erratum-ibid.\ B {\bf 600}, 453 (2001)]
  [Erratum-ibid.\ B {\bf 608}, 481 (2001)]
  [hep-lat/9911018].
  %%CITATION = HEP-LAT/9911018;%%
  %87 citations counted in INSPIRE as of 25 Aug 2014

%\cite{Luscher:1985wf}
\bibitem{Luscher:1985wf}
  M.~Luscher and P.~Weisz,
  %``Efficient Numerical Techniques For Perturbative Lattice Gauge Theory
  %Computations,''
  Nucl.\ Phys.\  B {\bf 266}, 309 (1986).
  %%CITATION = NUPHA,B266,309;%%

%\cite{DelDebbio:2008wb}
\bibitem{DelDebbio:2008wb} 
  L.~Del Debbio, M.~T.~Frandsen, H.~Panagopoulos and F.~Sannino,
  %``Higher representations on the lattice: Perturbative studies,''
  JHEP {\bf 0806}, 007 (2008)
  [arXiv:0802.0891 [hep-lat]].
  %%CITATION = ARXIV:0802.0891;%%
  %66 citations counted in INSPIRE as of 25 Aug 2014

%\cite{Weisz:1980pu}
\bibitem{Weisz:1980pu} 
  P.~Weisz,
  %``On the Connection Between the $\Lambda$ Parameters of Euclidean Lattice and Continuum {QCD},''
  Phys.\ Lett.\ B {\bf 100}, 331 (1981).
  %%CITATION = PHLTA,B100,331;%%

%\cite{Christou:1998ws}
\bibitem{Christou:1998ws} 
  C.~Christou, A.~Feo, H.~Panagopoulos and E.~Vicari,
  %``The Three loop Beta function of SU(N) lattice gauge theories with Wilson fermions,''
  Nucl.\ Phys.\ B {\bf 525}, 387 (1998)
  [Erratum-ibid.\ B {\bf 608}, 479 (2001)]
  [hep-lat/9801007].
  %%CITATION = HEP-LAT/9801007;%%

%\cite{Capitani:1994qn}
\bibitem{Capitani:1994qn} 
  S.~Capitani and G.~Rossi,
  %``Deep inelastic scattering in improved lattice QCD. 1. The First moment of structure functions,''
  Nucl.\ Phys.\ B {\bf 433}, 351 (1995)
  [hep-lat/9401014].
  %%CITATION = HEP-LAT/9401014;%%

   
\end{thebibliography}
\end{document}